\documentclass{sig-alternate-05-2015}

\usepackage{algorithm}
\usepackage{algpseudocode}
\usepackage[caption=false,font=footnotesize]{subfig}
\usepackage{url}

\usepackage{listings}
\usepackage[hidelinks]{hyperref}
\usepackage{times}
\usepackage{tabularx}

\usepackage{pgfplots, pgfplotstable}
\pgfplotsset{compat=newest}
\definecolor{carmine}{RGB}{150, 0, 24}
\definecolor{forest}{RGB}{34, 139, 34}
\definecolor{blues4}{RGB}{49, 130, 189}
\pgfplotsset{grid style={dashed,gray}}

\usepackage[normalem]{ulem}

\newif\ifsubmission
\newif\ifcomments
\newif\ifanonymous

\ifdefined\issubmission
\submissiontrue
\fi

\ifdefined\isanonymous
\anonymoustrue
\fi

\ifdefined\iscomments
\commentstrue
\fi

\ifsubmission
\else
\makeatletter
\def\@copyrightspace{\relax}
\makeatother
\fi

\ifcomments

\newcommand\remove[1]{\textcolor{red}{\sout{#1}}}
\else

\newcommand\remove[1]{}
\fi

\newcommand{\App}{$q$}
\newcommand{\Adv}{$\mathcal{ADV}$}
\newcommand{\KinibiTZ}{Kinibi-TZ}

\ccsdesc[500]{Security and privacy~Hardware-based security protocols}
\ccsdesc[300]{Security and privacy~Privacy-preserving protocols}

\begin{document}

\setcopyright{acmlicensed}
\CopyrightYear{2017}
\conferenceinfo{ASIA CCS '17,}{April 02 - 06, 2017, Abu Dhabi, United Arab Emirates}
\acmPrice{\$15.00}
\isbn{978-1-4503-4944-4/17/04}
\doi{http://dx.doi.org/10.1145/3052973.3053006}

\title{The Circle Game: Scalable Private Membership Test \\Using Trusted Hardware}


\numberofauthors{6}

\author{
Sandeep Tamrakar\\
       \affaddr{Aalto University}\\
       \email{\small{sandeep.tamrakar@aalto.fi}}
\alignauthor
Jian Liu\\
       \affaddr{Aalto University}\\
       \email{\small{jian.liu@aalto.fi}}
\alignauthor
Andrew Paverd\\
       \affaddr{Aalto University}\\
       \email{\small{andrew.paverd@ieee.org}}
\and
Jan-Erik Ekberg\thanks{This work was done while the author was at Trustonic.}\\
       \affaddr{Darkmatter}\\
       \email{\small{jan-erik.ekberg@darkmatter.ae}}
\alignauthor
Benny Pinkas\\
       \affaddr{Bar-Ilan University}\\
       \email{\small{benny@pinkas.net}}
\alignauthor
N. Asokan\\
       \affaddr{Aalto University \&\\ University of Helsinki}\\
       \email{\small{asokan@acm.org}}
} 

\maketitle

\ifsubmission
\else
\let\thefootnote\relax\footnotetext{\hspace{-1.5mm}This is the authors' extended version of the paper published at ACM ASIA CCS 2017.}
\let\thefootnote\relax\footnotetext{\hspace{-1.5mm}\url{http://dx.doi.org/10.1145/3052973.3053006}}
\fi

\subsection*{Abstract}
\label{sec:abstr}
Malware checking is changing from being a local service to a cloud-assisted one where users' devices query a cloud server, which hosts a dictionary of malware signatures, to check if particular applications are potentially malware. Whilst such an architecture gains all the benefits of cloud-based services, it opens up a major privacy concern since the cloud service can infer personal traits of the users based on the lists of applications queried by their devices. Private membership test (PMT) schemes can remove this privacy concern. However, known PMT schemes do not scale well to a large number of simultaneous users and high query arrival rates.      

We propose a simple PMT approach using a \emph{carousel}: circling the entire dictionary through trusted hardware on the cloud server.
Users communicate with the trusted hardware via secure channels. We show how the carousel approach, using different data structures to represent the dictionary, can be realized on two different commercial hardware security architectures (ARM TrustZone and Intel SGX).
We highlight subtle aspects of securely implementing seemingly simple PMT schemes on these architectures.
Through extensive experimental analysis, we show that for the malware checking scenario our carousel approach surprisingly outperforms Path ORAM on the same hardware by supporting a much higher query arrival rate while guaranteeing acceptable response latency for individual queries.

\keywords{ARM TrustZone; Intel SGX; Malware checking; Privacy}

\section{Introduction}
\label{sec:intro}
Malware checking used to operate primarily as a local service: a locally-installed anti-malware tool periodically receives lists of known threats from its vendor, but all its checks are done locally. 
This paradigm has already started to change in the era of cloud computing. 
Nowadays, anti-malware tools are often thin clients and the bulk of the threat data is held by a cloud-based service. 
The anti-malware tool consults the cloud service to determine if a particular application is likely to be malware. 
Such a design pattern is attractive for anti-malware vendors for a variety of reasons: it avoids unnecessary data transfers, ensures that all users have up-to-date threat information, and allows the anti-malware vendor to retain its full set of known malware signatures as a potential competitive advantage without having to disclose it in full to all customers.
For example, Google's \emph{Verify Apps} is a cloud-based service that checks if a mobile application is potentially harmful before it is installed~\cite{Android_Sec_Report}.
Protocols involving such remote lookup operations also occur in other scenarios, such as querying whether a document contains a malicious payload, or checking if a password is present in a database of leaked passwords.
Abstractly, such a lookup operation is a remote \emph{membership test}: a user holding an item $q$ wants to check if $q$ is a member of a large set $X$, called the \emph{dictionary}, held by a remote server.

Although services built on remote membership test have significant advantages, they suffer from a major privacy concern -- the server learns all queries submitted by users, which allows inference of sensitive personal information.
For example, it has been demonstrated that the set of applications on a user's device can be used to infer the user's gender, age, religion, and relationship status~\cite{Seneviratne2014}. 
Service providers who want to use such remote membership tests therefore wish to demonstrably preclude any ability to infer personal information about their users~\ifanonymous\cite{FSC-anon}\else\cite{FSC}\fi. 
The natural cryptographic primitive to build such a \emph{private membership test (PMT)} is \emph{private set intersection (PSI)}~\cite{Pinkas2014, Pinkas2015}. 
However, PSI schemes have two major drawbacks: high complexity and poor scalability.
First, a PSI involving a dictionary of size $n$, requires $\mathcal{O}(n)$ communication between client and server and requires the server to perform $\mathcal{O}(n)$ operations.
Second, PMT servers receive a large number of simultaneous queries, but PSI-based schemes are not amenable to aggregation of queries.

Trusted hardware has now become widely available on commodity computing platforms. 
Trusted execution environments (TEEs) are already pervasive on mobile platforms~\cite{EKA14}, and newer TEEs such as Intel's SGX~\cite{McKeen2013} are being deployed on PCs and servers.
Several prior works~\cite{
Backes2012, Iliev2005, Jacob2013}
show how trusted hardware can be used to establish a \emph{trust anchor}~\cite{Schiffman2010} in the cloud. 
The combination of such a trust anchor with Path ORAM~\cite{DBLP:conf/ccs/StefanovDSFRYD13}, the recent breakthrough in oblivious random access memory (ORAM), can be used to solve the PMT problem. 
This solution has constant communication overhead and only $\mathcal{O}(\log n)$ computational overhead per query. 
However, like PSI, Path ORAM is not amenable to aggregation of simultaneous queries. 
Therefore, supporting $m$ simultaneous queries will incur $\mathcal{O}(m \log n)$ cost.



However, in cloud-based mobile malware checking, it is critical to optimize for number of queries performed ($m$) rather than dictionary size ($n$). 
To illustrate this: in 2015, anti-malware vendor Kaspersky identified nearly 3~million malicious mobile installation packages~\cite{Kaspersky_mobile_malware}.
In contrast, Google reported that at the end of 2015 over 1~billion devices were protected by its end-point security services, and that it performed over 400~million device scans per day~\cite{Android_Sec_Report}.
Furthermore, it is estimated that each Android device has on average 95 apps installed~\cite{YahooAviate}.
Therefore, the $\mathcal{O}(m \log n)$ cost incurred by Path ORAM will be too high -- the ideal PMT service for this scenario should maximize the throughput of queries.

For this use case, the time required to respond to a query (i.e. the \emph{response latency}) must be within a couple of seconds.
More importantly, the service must be able to guarantee an upper bound on its response latency, even as the arrival rate of queries increases.
If a PMT scheme provides this guarantee for a given query arrival rate, we say the scheme is \emph{sustainable} at that rate. 
For every scheme there is a maximum query arrival rate beyond which the response latency will keep increasing over time.
We call this the \emph{breakdown point}.
At this point, new hardware must be added to the server in order to guarantee response latency.

Another important characteristic of cloud-based mobile malware checking is that it is acceptable for the PMT protocol to exhibit a small but non-zero false-positive rate (FPR).
A user who receives a positive result from the service can resubmit the same query directly to the anti-malware vendor to ascertain the true result and receive guidance on how to remediate the potential threat.
If the FPR is sufficiently low, only a small fraction of apps will be revealed to the dictionary provider, thus still protecting users' privacy.

Based on these characteristics, we propose an effective and efficient PMT scheme that can support a significantly larger number of simultaneous queries compared to known PMT schemes.
Our contributions are as follows:

\begin{itemize}
  \itemsep0em

\item We introduce a new \textbf{carousel design pattern} in which the dictionary (or a representation thereof) is continuously circled through trusted hardware on the lookup server (Section~\ref{sec:carouselApproach}). 
This prevents the lookup server from learning the contents of the queries, and also guarantees low query response latency.

\item We show how the system's performance can be significantly improved by selecting efficient \textbf{data structures to represent the dictionary}.
We evaluate several different data structures (Section~\ref{sec:dictRep}), and describe how to construct and process each without leaking information.

\item Through a systematic and extensive experimental evaluation using two different commercial hardware security architectures (ARM TrustZone and Intel SGX), we show that for typical parameters in the malware checking scenario, our carousel-based PMT can support a \textbf{large number of simultaneous queries} while still guaranteeing sufficiently low response latency for every query (Section~\ref{subsec:experimentdesign:impl}). 


\item We also describe how to solve the PMT problem using Path ORAM (Section~\ref{subsec:experimentdesign:CoO}) and experimentally compare this against our carousel approach. 
Although the ORAM-based scheme achieves very low query response latency, it reaches its breakdown point quickly. 
In contrast, the carousel approach provides a more modest query response latency while sustaining \textbf{much higher query arrival rates with the same hardware} -- 2.75 times on Intel SGX and nearly 10 times on ARM TrustZone (Section~\ref{subsec:experimentdesign:eval}).


  
\end{itemize}


\pagebreak

\section{Preliminaries}
\label{sec:preliminaries}

\subsection{Trusted Execution Environment}
\label{sub:sec:tee}



A \emph{Trusted Execution Environment} (TEE) is a system security primitive that isolates and protects security-critical logic from all other software on the platform.
All software outside the TEE is said to be running in the \emph{Rich Execution Environment} (REE), which usually includes the operating system and the majority of the platform's software.
A piece of application logic running in the TEE is referred to as a \emph{Trusted Application} (TA), whilst an application running in the REE is a \emph{Client Application} (CA).
Fundamentally, a TEE protects the confidentiality and integrity of a TA's data, and ensures that no REE software can interfere with the TA's operation.
A TEE usually provides some form of \emph{remote attestation}, which allows remote users to ascertain the current configuration and behavior of a TA.
The combination of these capabilities enables remote users to trust a TA.
In modern systems, the capability to establish and enforce a TEE is often provided by the CPU itself.
This leads to very strong hardware-enforced security guarantees, and also improves performance by enabling the TEE to execute on the main CPU.
However, in some cases this may allow malicious software in the REE to mount side-channel attacks against the TEE.
Our design does not depend on any platform-specific features, and can thus be realized on any TEE that exhibits the above characteristics.
We demonstrate this by implementing our system on the two most prevalent commercial TEEs: ARM TrustZone and Intel SGX.

\subsubsection{ARM TrustZone}
\label{sub:sub:tz}


ARM TrustZone\footnote{\url{https://www.arm.com/products/security-on-arm/trustzone}} is a contemporary TEE architecture that is widely deployed on smartphones and is now being deployed on infrastructure-class AMD CPUs.\footnote{\url{http://www.amd.com/en-us/innovations/software-technologies/security}}
TrustZone provides a platform-wide TEE, called the \emph{secure world}, which is fully isolated from the REE or \emph{normal world}.
All interaction between the REE and TEE is mediated by the CPU.
In order to support multiple TAs, the secure world usually runs a trusted OS, such as Kinibi from Trustonic.\footnote{\url{https://www.trustonic.com/products/kinibi}}
Due to the constraints of the platform, the trusted OS may limit TA's internal code and data memory (e.g. Kinibi limits each TA to 1~MB).
The platform can be configured to map TA's internal memory to system-on-chip (SoC) RAM.
With such a configuration, TA's internal memory is \emph{secure memory} since TrustZone protects its confidentiality and integrity against an adversary who controls the normal world.
Furthermore, TA's internal memory is \emph{private memory} since the adversary cannot observe TA's \emph{memory access pattern} (i.e. the metadata about which addresses are being accessed, and in what order).
In contrast, the adversary can observe all accesses TA makes to the REE memory.

In Kinibi on ARM TrustZone ({\em \KinibiTZ{}}), interaction between a normal world CA and a TA in the TEE follows a request-response pattern: CA can invoke a specific operation provided by TA.
In addition to a small set of TA invocation parameters, CA can usually demarcate up to 1~MB of its memory to be shared with TA.
In the same way that memory can be shared between applications on any modern OS, the memory management unit (MMU) maps a physical memory page to the virtual address spaces of both CA and TA.
This page can be read and written by both endpoints, and the processor's mechanisms for cache coherency ensure that memory accesses are properly synchronized.
This feature allows TA to access large portions of normal world memory.





\subsubsection{Intel SGX}
\label{sub:sub:sgx}

Intel's recent \emph{Software Guard Extensions} (SGX) technology~\cite{McKeen2013} allows individual applications to establish their own TEEs, called \emph{enclaves}.
An enclave can contain application logic and secret data, protecting the confidentially and integrity of them from all other software on the platform, including other enclaves, applications, or the (untrusted) OS.
SGX includes remote attestation capabilities to provide remote parties with assurance about the code running in an enclave~\cite{Anati2013}.
For consistency, we refer to the untrusted application that hosts the enclave as \emph{Client Application} (CA), and the enclave itself as \emph{Trusted Application} (TA).
Although both SGX and TrustZone have similar objectives, the specific architectures of these two technologies give rise to several important differences.

\noindent\textbf{Memory considerations.}
Unlike TrustZone's platform-wide TEE, SGX supports multiple enclaves: each TA runs in its own enclave.
Each enclave is part of an application and runs in the same virtual address space as its host application.
This means that the enclave can directly access the application's memory, but attempts by the application or OS to access the enclave memory are blocked by the CPU.
Whenever any of the enclave memory leaves the CPU (e.g. is written to DRAM), it is automatically encrypted and integrity-protected by the CPU.
However, even though SGX provides \emph{secure memory} (i.e. confidential and integrity-protected), the enclave's memory access pattern may still be observable by untrusted software on the same platform.
This lack of \emph{private memory} potentially gives rise to the following classes of side-channel attacks:

\noindent\textbf{Deterministic side-channel attacks.}
Xu et al.~\cite{Xu2015} have shown how a malicious OS can manipulate the platform's global memory page table, which includes the enclave memory pages, to cause page faults whenever the enclave reads from or writes to its memory.
If the enclave's memory access pattern depends on some secret data, their technique can be used to discover its value by observing the sequence of page faults.
This side-channel attack is deterministic and thus can be effective even with only a single execution trace.
However, the adversary can only observe memory accesses at page-level granularity (usually 4~kB).
For example, he can observe 
when a particular memory page is accessed and can distinguish between reads and writes, but cannot ascertain the specific addresses of these operations within the page.

\noindent\textbf{Probabilistic side-channel attacks.}
Liu et al.~\cite{Liu2015b} have presented an even stronger cache side-channel attack, which could be used against SGX.
They exploit the fact that the CPU's level~3 (L3) cache is shared between all cores, and that the adversary may have control of the other cores while the enclave is executing.
Through this type of attack, an adversary may be able to observe the enclave's memory access pattern at cache line (CL) granularity (usually 64~B).
However, since the adversary does not have direct control of the L3 cache, this is a probabilistic attack that requires the \emph{secret-dependent memory accesses to be repeated multiple times}.

In this paper, we assume that SGX enclave memory can be considered private at page-level granularity.
That is, different accesses \emph{within a page of enclave memory} are indistinguishable to an adversary. 
Accesses to different enclave pages can be noticed by the adversary, even though the pages' contents are encrypted and integrity-protected.
Therefore, as explained in Section~\ref{sec:experimental_evaluation}, we ensure that in all our SGX implementations the page-level memory access patterns do not depend on secret data. 
On the other hand, probabilistic cache side-channel attacks are generally infeasible if secret-dependent memory accesses are not repeated multiple times.
We therefore ensure that none of our SGX implementations perform secret-dependent memory accesses more than once. 
If stronger resistance to probabilistic side-channel attacks is required, techniques such as those used in \emph{Sanctum}~\cite{Costan2016} could be applied.

\subsection{Oblivious RAM}
\label{sub:sec:oram}

{\em Oblivious RAM} (ORAM) is a cryptographic primitive originally proposed by
Goldreich and Ostrovsky~\cite{Goldreich1996} to prevent information
leakage through memory access patterns.  In ORAM schemes, a secure processor (e.g. TEE) 
divides its data into blocks, which it encrypts and stores in randomized order in non-secure 
memory, such as the platform's main memory. On each access, the processor reads the
desired block and some dummy data, and then re-encrypts and reshuffles this data before writing it back to non-secure memory. The processor also needs to update some state
in its private memory. Under ORAM, every access pattern is
computationally indistinguishable from other access patterns of the same
length.

The state-of-the-art ORAM techniques are tree-based
constructions~\cite{Ren2015, DBLP:conf/ccs/StefanovDSFRYD13, Devadas2016, Dachman-Soled2015, Boyle2016},
where the data blocks are stored in a tree structure. 
For example, in Path ORAM the processor stores a {\em position map} in its
private memory to record the path in which each block resides. When the processor
wants to access a block, it reads the block's complete path from the root to the
leaf. To store and access $n$ blocks from insecure memory, tree-based ORAM
has a bandwidth cost of $\mathcal{O}(\log n)$ and uses $\mathcal{O}(\log n)$
private memory (if recursively storing the position map).



\section{Problem Setting}
\label{sub:systemodel}

\begin{figure}[t]
\centering
\centerline{\includegraphics[scale=0.34,trim={0 12mm 0 12mm},clip]{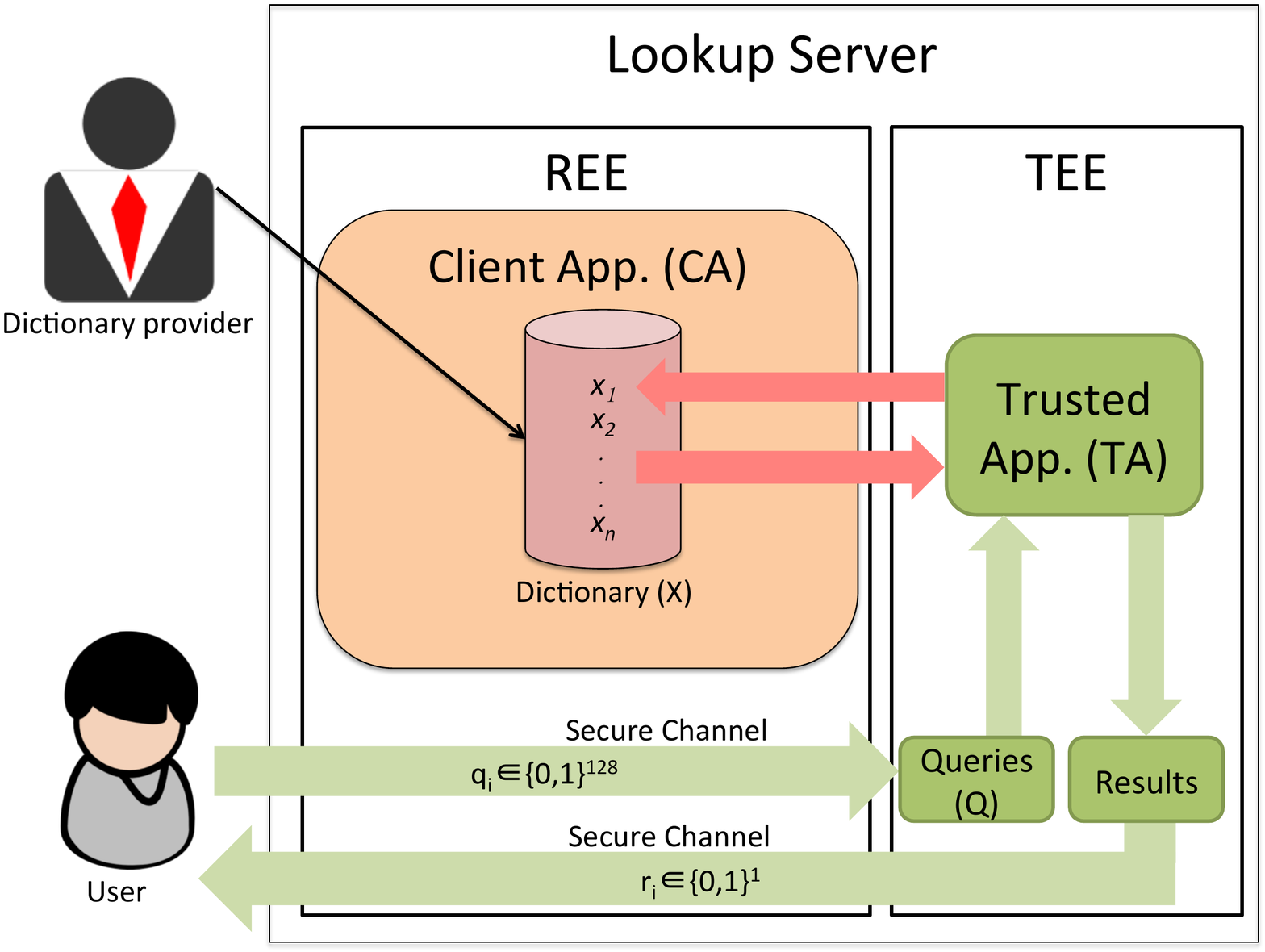}}
\caption{System model: cloud-based private membership test.}
\label{fig:legacy}
\end{figure}


\subsection{System Model}
\label{sec:system-model}
Figure~\ref{fig:legacy} depicts a generalized system model for cloud-based private membership test (PMT) using trusted hardware. 
It consists of a \emph{dictionary provider}, a \emph{lookup server}, and \emph{users}.
We describe and evaluate our system in terms of the concrete use case of cloud-based malware checking, but we emphasize that our approach can be applied to many other use cases. 

In the cloud-based malware checking scenario, the \emph{dictionary provider} is the anti-malware vendor that constructs and maintains a malware dictionary $X=\{x_1, ..., x_n\}$ containing $n$ entries. 
Each entry $x_i$ in $X$ is a unique malware identifier.
The lookup server is a remote server that provides malware checking functionality to users.
The lookup server could be operated by a third party, such as a content delivery network. 
The actual lookup functionality is provided by a TA running in the TEE.
The lookup server also runs a CA in its REE, which facilitates interaction between users and TA, and makes $X$ accessible to TA.
In general, the dictionary may grow arbitrarily large and thus cannot be stored inside the TEE.

Each user can authenticate and attest TA before establishing a secure communication channel with TA.
The user then submits a query (\App{}) representing an application.
TA stores the received queries $Q=\{q_1, ..., q_m\}$ in its secure memory, where $m$ is the number of concurrent queries at any given time. 
TA must return a single bit: \texttt{1} if \App{} $\in X$ and \texttt{0} otherwise.
These responses are also kept in secure memory until they are ready to be returned to the users via their respective secure channels.
The primary privacy requirement is that the adversary (with capabilities defined in Section~\ref{sub:adversary}) must not learn any information about \App{}.
Note that the adversary is permitted to learn statistical information such as the number of queries submitted by a particular user (e.g. through traffic analysis) or the total number of queries currently being processed.
Hiding the communication patterns between users and the lookup server is an orthogonal problem.

\subsection{Mobile Malware Use Case Parameters}

As explained in Section~\ref{sec:intro}, in 2015 approximately 3~million mobile malware samples were detected -- an increase of nearly 900~000 over the previous year~\cite{Kaspersky_mobile_malware}.
Therefore, we target a malware dictionary of $2^{26}$ entries (\textasciitilde 67 million entries) as a reasonable estimate for the next ten years.
Furthermore, in the cloud-based mobile malware checking scenario, it is acceptable for the PMT protocol to exhibit a small but non-zero false-positive rate (FPR) without significantly diminishing users' privacy.
However, false negatives are never permissible.
This tolerance of a non-zero FPR can provide significant performance benefits, as we show in Sections~\ref{sec:dictRep} and \ref{sec:experimental_evaluation}.
Based on the recommendation of a leading anti-malware vendor, we selected an FPR of $2^{-10}$ which implies that the majority of users will encounter at most one false positive in total~\ifanonymous\cite{FSC-anon}\else\cite{FSC}\fi.
Given that the average Android user has 95 installed apps~\cite{YahooAviate}, the majority of users will encounter at most one false positive in total.



\subsection{Adversary Model}
\label{sub:adversary}
The primary adversary we consider is a malicious lookup server, which is assumed to have full control of the REE.  
Its objective is to learn information about the contents of users' queries, which can be used to profile users.
As usual, we assume that the adversary is computationally bounded and cannot subvert correctly implemented cryptographic primitives.  
Therefore the secure channels between TA and users prevent the lookup server learning the content of messages exchanged via these channels.
Furthermore, we assume the adversary will not perform hardware-level attacks due to the relatively high cost of such attacks compared to the value of the data.
Therefore, the adversary cannot observe or modify the internal state of TA or TA's interactions with its data in private memory.
On the other hand, a lookup server can masquerade as a user and submit its own arbitrary queries.
It can also schedule or remove incoming queries as it sees fit.
It can observe and measure the duration of TA's interactions with non-private memory and individual query response latencies.
It can attempt to use any available software side channel (deterministic or probabilistic), and it may attempt to modify or disclose the dictionary.

Our secondary adversary is the dictionary provider itself, which is assumed to be honest-but-curious.
It may attempt to infer information about users or profile them based on any application identifiers revealed to it by the users.
However, it is assumed to only add legitimate malware identifiers to the dictionary ($X$).
Enforcing this behaviour is an orthogonal problem, which may be addressed by e.g. introducing reputation scores for dictionary providers.
The dictionary provider authenticates $X$ towards TA (e.g. via a message authentication code using a key it shares with TA) so that TA can detect any tampering of $X$.
If the dictionary provider wants to keep the dictionary confidential from the lookup server, it is also possible to encrypt the dictionary such that it can only be decrypted by TA.
We deem denial-of-service attacks to be out-of-scope.

\section{Requirements} 
We define the following requirements to ensure the system's security, performance and accuracy: 
\begin{enumerate}
\renewcommand{\theenumi}{R\arabic{enumi}}
\renewcommand{\labelenumi}{\theenumi.}
\itemsep0em
\item \label{R1}\textbf{Query Privacy}: The lookup server and dictionary provider must not be able to learn anything about the content of the users' queries or the corresponding  responses.
The dictionary provider may learn the content of queries for which TA gave a positive response (i.e. potentially malicious applications), if the user chooses to reveal these.
Stated in the ideal/real model paradigm: If there were an inherently trusted entity (ideal model), then it could have received the dictionary from the dictionary provider and the queries from users, and sent responses to users without leaking any other information to any entities. 
We require that a real world solution does not disclose more information than this ideal model.
\item \label{R2}\textbf{Response latency}: The service must answer every query in an acceptable time (e.g. in 2 seconds for the malware checking use case). This response latency must be sustainable.
\item \label{R3}\textbf{Server scalability}: The service must be able to sustain a level of overall throughput (i.e. queries processed per second) that is sufficient for the intended use case (e.g. in the order of thousands of queries per second for the malware checking use case).
\item \label{R4}\textbf{Accuracy}: The service must never respond with a false negative.
The false positive rate must be within the acceptable limits for the intended use case. 
\end{enumerate}


\section{The Carousel approach}
\label{sec:carouselApproach}

\begin{figure}
\centering
\centerline{\includegraphics[scale=0.34,trim={0 12mm 0 12mm},clip]{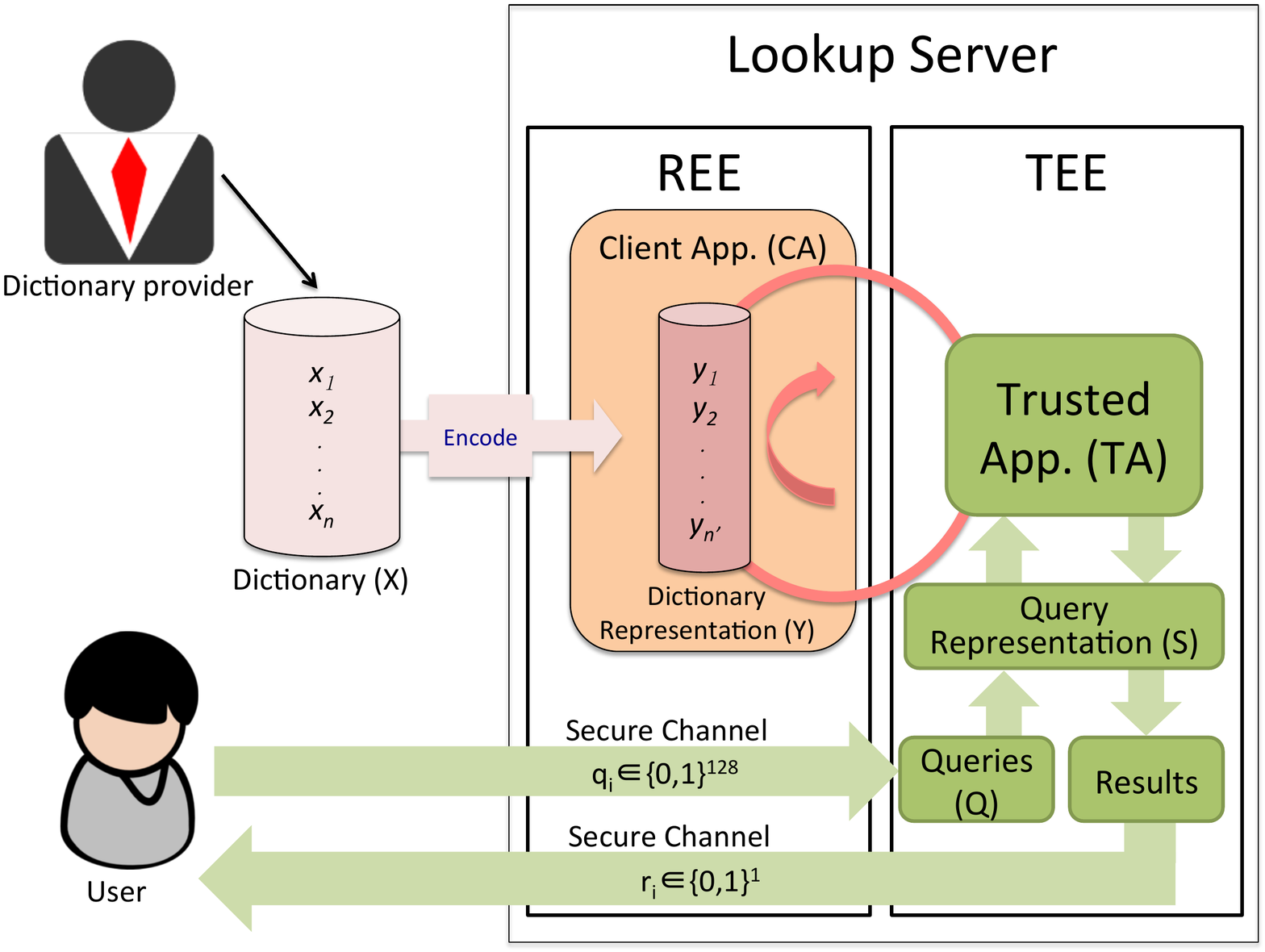}}
\caption{Overview of the carousel approach.}
\label{fig:carousel}
\end{figure}

To meet the requirements defined in the preceding section, TA needs a mechanism for accessing the dictionary ($X$) without leaking any information about the users' queries (Requirement \ref{R1}).
The naive approach of accessing specific elements of $X$ in the REE violates this requirement because the adversary can observe which dictionary items are being checked.
Canonically, this type of problem could be solved using ORAM where TA is the ORAM processor and REE stores the encrypted shuffled database.
However, as an alternative approach, we propose a new \emph{carousel design pattern} in which a representation of the dictionary is continuously circled through TA.
As we demonstrate in Section~\ref{sec:experimental_evaluation}, the fundamental advantage of our carousel approach is that it supports efficient processing of \emph{batches of queries} in a single carousel cycle.
Namely, whereas using ORAM to answer a batch of $m$ queries requires accessing $\mathcal{O}(m\log n)$ dictionary items, these $m$ queries can be answered by a single cycle that reads $n$ dictionary items.
When the size of the batch increases, the latter approach becomes more efficient.
Figure~\ref{fig:carousel} gives an overview of our carousel approach.

\noindent\textbf{Dictionary Representation.}
To avoid leaking information about a query's position in the dictionary ($X$), TA cannot send a response until it has completed one full carousel cycle, starting from the point at which the query arrived.
Therefore, the two factors influencing response latency are the size of the dictionary and the processing efficiency.
To minimize latency, the Dictionary Provider transforms the dictionary $X$ into a more compact and/or more efficient data structure, which we call the \emph{dictionary representation} $Y=\{y_1, ..., y_{n^\prime}\}$, which is cycled through TA.
The choice of data structure therefore has a significant impact on the performance of the system, and although several well-known data structures support efficient membership tests, it is not obvious which is best-suited for the carousel setting. 
In Section~\ref{sec:dictRep} we discuss these different data structures and in Section~\ref{sec:experimental_evaluation} we experimentally evaluate their performance.


\noindent\textbf{Query representation.}
TA transforms queries $Q$ into representations $S=\{s_1, ...,s_{m^\prime} \}$ in a similar manner to $Y$ and maintains them in sorted order.
TA stores $Q$ in its secure memory along with the queries' times of arrival and references to their representations in $S$.
When more than one query maps to the same representation $s_k$, TA only maintains a single $s_k$ in $S$, but adds dummy query representations in $S$ and keeps track of the number of queries referencing $s_k$.
Keeping a single instance of $s_k$ in $S$ irrespective of the number of queries that maps to it allows the TA to only operate on the single $s_k$.
This is required to prevent information leakage, since the adversary is also permitted to submit queries. 

\noindent\textbf{Carousel processing.}
To process queries, TA cycles through $Y$ and scans its contents in order to answer the received queries.
CA divides $Y$ into several chunks and invokes TA sequentially with each chunk as input along with waiting queries. 
We assume that queries arrive continuously and breaking $Y$ into chunks allows queries to be passed to TA without having to wait for a full carousel cycle.
Incoming queries are associated with the identifier of the chunk with which they arrived, which is defined as their time of arrival.
At the beginning of each chunk, TA updates $S$ based on the newly received queries.  
TA then compares each entry in the chunk with $S$ and records the results. This process is repeated for each chunk.

\noindent\textbf{Response construction.} 
When a query has waited for a full cycle of $Y$, TA processes the accumulated results and computes the response.
Responses are sent to the users at the end of each invocation of TA.
Once a response has been sent, TA removes the query from $Q$ and removes its representations from $S$ if there are no other queries associated with those representations, otherwise it removes dummy representations from $S$.

\noindent\textbf{Avoiding information leakage.}  
As explained in Section~\ref{sub:adversary}, the adversary \Adv{} can observe memory access patterns for all non-private memory (including CA's memory), and can measure the time taken to respond to each query.
To provide the strongest possible security guarantees, we assume that \Adv{} knows exactly which entry in $Y$ is currently being processed by TA.
If \Adv{} could determine whether or not this entry is relevant to the current set of queries, which in the worst case could be a single query, this would leak information.
Furthermore, for a given query, it is possible that TA could respond before completing a full carousel cycle (e.g. if the relevant information was found at the start of $Y$).
However, since $Y$ is not secret and \Adv{} knows which chunk is currently being processed by TA, the time between query arrival and response might also leak information about both the query and response.
Therefore, in the carousel approach, we can satisfy Requirement~\ref{R1} by a) performing constant-time processing for every entry in $Y$, and b) ensuring that every query remains in TA for exactly one full carousel cycle.
In other words, the number of operations TA performs per chunk must be independent of $S$, and the query response latency must be independent of the queried value.
\section{Dictionary Representation}
\label{sec:dictRep}

As explained in the preceding section, the performance of the system can be significantly improved by choosing an efficient data structure with which to represent the dictionary.
Although there are various data structures that support efficient membership tests in general (e.g. Bloom filter~\cite{Bloom1970}), it is not obvious which of these is best-suited for use in the carousel approach.
Since query latency depends on the length of the dictionary and the cost of processing each entry, the ideal dictionary representation would minimize both of these aspects (Requirement~\ref{R2}).
Furthermore, the chosen data structure must support efficient batch processing (i.e. answering multiple queries in each carousel cycle), since this is the fundamental advantage of the carousel approach and also improves server scalability (Requirement~\ref{R3}).
In this section we explore different data structures for representing the dictionary.
We first discuss the naive approach of using an unmodified dictionary, but show that this is always less space-efficient than our new \emph{Sequence of Differences} representation in which we encode the differences between successive dictionary entries.
We then describe how to use two well-known data structures, \emph{Bloom filter} and \emph{4-ary Cuckoo hash}, in the carousel setting.
Finally, we compare the size and processing complexity of these different representations.

As explained in Section~\ref{sec:intro}, our motivating scenario of cloud-based malware checking can tolerate a low but non-zero false positive rate (FPR).
We argue that this is also a reasonable assumption for other such applications of a PMT protocol.
This is important because it enables us to use data structures with an inherently non-zero FPR (e.g. Bloom filter) or to reduce the size of the dictionary representation (e.g. using shorter hashes in the 4-ary Cuckoo hash representation).
We denote the acceptable FPR as $2^{-\varepsilon}$ and explain how this is determined for each representation.


\subsection{Naive Approach}

The most naive approach is to cycle the unmodified dictionary entries through TA (i.e. $Y = X$) and compare these against the queries.
This is suboptimal because the dictionary entries could be arbitrarily large, thus increasing the size of $Y$ unnecessarily.

Given that it is acceptable to have an FPR of $2^{-\varepsilon}$, a slightly better naive method is to map each dictionary entry $x_i$ uniformly to a point in a domain of size $n \cdot 2^{\varepsilon}$. 
The FPR can be calculated as: 
\begin{center} 
FPR = $1-(1-1/(n\cdot2^{\varepsilon}))^{n}\approx 1- e^{-1/2^{\varepsilon}}\approx 1/2^{\varepsilon} = 2^{{-\varepsilon}}$,
\end{center}
Therefore $(\varepsilon+\log n)$ bits are needed in order to represent an item, and thus the length of $Y$ is $n\cdot(\varepsilon + \log n)$ bits.
The same mapping is applied to the queries such that the resulting query representations can be compared against $Y$.
However, this approach always results in a larger $Y$ compared to our new Sequence of Differences representation, as described in the following subsection.
We therefore elide the naive approach from our comparisons and use the Sequence of Differences representation as our baseline.


\subsection{Sequence of Differences} 
\label{sub:diffSeq}
 
\noindent\textbf{Dictionary representation.}
Compared to the naive approach, we can reduce the size of $Y$ by representing only the differences between successive items, with minimal additional processing cost.
We first hash each entry $x_i$ to a value $h_i$ of length $(\varepsilon + \log n)$, and sort the
resulting values: $h_{0}<h_{1}<\cdots<h_{n}$. Alternatively, $h_i$ can simply be a truncation of $x_i$, since the entries are already
uniformly distributed in the malware checking case.
Instead of storing the entries themselves in $Y$, we only store the differences between successive entries: $y_0 = h_{0}, y_1 = h_{1}-h_{0},\ldots,y_n = h_{n}-h_{n-1}$.
If multiple entries result in the same $h_j$, we only keep one copy of $h_j$ in $Y$ to avoid leaking information.
%
The advantage of this approach is that the length of the differences ($y_i$ values) is smaller than the length of items ($h_i$ values).
However, this approach requires choosing a fixed number of bits to represent all differences. 

We ran a simulation which showed that the probability of a difference being larger than
$(2^{{\varepsilon+2}}-1)$ is approximately 2\%.  Therefore, we chose to use $(\varepsilon+2)$ bits
to represent a single difference.  
In the vast majority of cases, the difference $y_{i}= h_{i}-h_{i-1}$ is less than $2^{{\varepsilon+2}}$, so we insert it directly into $Y$.
Otherwise, $y_i = p\cdot(2^{{\varepsilon+2}}-1)+b$, where $b < 2^{{\varepsilon+2}}-1$.
In this case we insert $p$ entries of ``zero'' (each $\varepsilon+2$
bits) into $Y$, followed by $b$ (with $\varepsilon+2$ bits as usual).
Note that since the actual difference $y_{i}$ is always greater than $0$, it is easy to recognize these dummy entries.
We expect to add about $0.02n$ dummy entries, so the total size of $Y$ remains approximately $1.02(\varepsilon+2)n$.

\noindent\textbf{Query representation.} 
TA maps each query to its representation in $S$ by applying the same operation as for $x_i$. TA maintains $S$ as a sorted list with $m$ unique items, each $(\varepsilon + \log n)$ bits in length. 

\noindent\textbf{Carousel processing.} 
Algorithm~\ref{algo:difference} shows the carousel processing for a chunk of $Y$. 
TA first recovers the value of the current dictionary entry $h_i$ by adding the difference $y_i$ to the previous entry $h_{i-1}$.
For each recovered entry $h_i$, it uses binary search to check and mark whether $h_i$ is in $S$. 
The time taken for this binary search must not depend on the values of the current queries, and thus TA spends equal time processing every $h_i$.
If TA stopped after finding $s_j \leq h_{i}$, an adversary \Adv{}, who knows $Y$, could measure response latency to learn whether a certain query is in $S$ (note: \Adv{} can also insert false queries to influence response latency).
%
Overall, it takes $\mathcal{O}(n\lceil\log m\rceil)$ operations to process each $h_i$.
Whenever $y_i = 0$, TA identifies this as a dummy item and adds $(2^{{\varepsilon+2}}-1)$ to $h_{i-1}$, but it continues without performing a binary search for the $h_i$, since $Y$ is already known to the adversary.
This algorithm ensures that TA spends equal time for non-zero entries in $Y$.

\noindent\textbf{Response construction.} 
When a query completes one carousel cycle, TA generates its response by checking if the corresponding item in $S$ is marked as a match.

\begin{algorithm}[ht]
\caption{Membership test using \emph{Sequence of Differences}}
\label{algo:difference}
\begin{algorithmic}
    \State $Y$: Dictionary representation
    \State $S$: Query representation
    \State $h$: Current entry
     \State $i = 0$
    \While{$i$ is in the current chunk}
      \If{$Y[i]$ equals 0}
      	\State $h \gets h + 2^{\varepsilon+2} - 1$
      \Else
      	\State $h \gets h + Y[i]$
      	\State $binary\_search$ of $Y[i]$ in $S$ 
      \EndIf
      \State $i++$
    \EndWhile
\end{algorithmic}
\end{algorithm}

\subsection{Bloom Filter}

\noindent\textbf{Dictionary representation.} 
A {\em Bloom filter} is a data structure used for efficient membership testing.  It is an $N$-bit array $B$
initialized with $0$s, together with $l$ independent hash functions $H_i(\cdot)$ whose output is uniformly
distributed over $[0, N-1]$~\cite{Bloom1970}.  To add an entry $x$ to the filter, we compute $l$ array
positions: $h_i=H_i(x)$,  $\forall i, 1 \leq i \leq l$, and set each of these $l$ positions in $B$ to 1
($B[h_i]=1$).  To test if an item is in the dictionary, $l$ positions are calculated using the same set of
hash functions. If any of these positions in $B$ is set to $0$, we can conclude that the item is not in
$B$. Otherwise, the item is declared to be  in $B$. 
The false positive rate is:
\begin{center}
FPR$_{bf} = (1-(1-\frac{1}{m})^{nl})^l \approx (1-e^{-\frac{nl}{m}})^{l}$,
\label{fpbf}
\end{center}

For an FPR of $2^{{-\varepsilon}}$, an optimized bloom filter needs $1.44\varepsilon n$ bits to store $n$ items~\cite{Pagh2005}. We represent the Bloom filter as a bit array $Y$, which is the dictionary representation.

\noindent\textbf{Query representation.} 
For each query, TA calculates $l$ byte positions in the bloom filter
and adds the positions to $S$ in sorted order. 

\noindent\textbf{Carousel processing.} 
Algorithm~\ref{algo:bf} shows the carousel processing for a chunk of $Y$. The algorithm essentially copies from the carousel all bytes containing data required to decide whether the queries are in the dictionary (namely, the bytes to which the queries are mapped by the hash functions).
$R$ is a list of bytes for storing results, initialized to zeros.
For each byte in the current chunk, TA checks whether the byte is needed, as indicated by $S$. 
If so, it copies the byte to $R$. Otherwise, it copies the byte to a dummy location $dummy\_byte$. 
TA performs an equal number of operations for every byte in $Y$.

\noindent\textbf{Response construction.} Once the carousel processing completes, TA goes through all the queries, links them back to the query representation, and inspects the corresponding values in $R$ to check if all bit positions for a particular query are set.

\begin{algorithm} [ht]
\caption{Membership test using \emph{Bloom Filter}}
\label{algo:bf}
\begin{algorithmic}
    \State $Y$: Dictionary representation
    \State $S$: Query representation
    \State $R$: A list empty bytes
    \State $dummy\_byte$: a byte used to do dummy operations 
    \State $dummy\_int$: an integer used to do dummy operations 
    \State $i = 0, \quad j = 0$
    \While{$i$ is in the current chunk}
      \If {$S[j]$ equals $i$}
        \State $R[j] \gets Y[i]$
        \State $j++$ 
      \Else       
        \State $dummy\_byte \gets Y[i]$
        \State $dummy\_int++ $
      \EndIf
      \State $i++$           
    \EndWhile
\end{algorithmic}
\end{algorithm}


\subsection{4-ary Cuckoo hash}
\label{sub:cuckoo}
{\em Cuckoo hash} is another data structure for efficient membership test~\cite{Pagh2004}.
We use a variant, called d-ary Cuckoo hash with four hash functions, since it utilizes approximately 97\% of the hash table (compared to less than 50\% utilization in standard Cuckoo hash with two hash
functions)~\cite{Erlingsson06}.

\noindent\textbf{Dictionary representation.} 
Four hash functions $H_1$ -- $H_4$ are used to obtain four candidate positions for a given dictionary
entry $x_i$ in $Y$.  During insertion, $x_i$ is hashed to a ($\varepsilon$+2)-bit value $y_i$ (e.g. by
truncating $x_i$). This value is inserted into the first available candidate position.  If all 4 positions for a
given $y_i$ are already occupied (say, by values $\hat{y_1}, \hat{y_2}, \hat{y_3}, \hat{y_4}$), $y_i$ is inserted by
recursively relocating $\hat{y_j}$ into one of their 3 other positions (since each $\hat{y_j}$ has a choice of 4
positions in $Y$).  In the worst case, this recursive strategy could take many relocations or get into an
infinite loop. 
The standard solution for this problem is to perform a full rehash, but the probability for this event is
shown to be very small.
Kirsch et al.~\cite{Kirsch2010} introduce a very small constant-sized auxiliary stash for putting the current unplaced item when a failure occurs. 
They show that this can dramatically reduce the insertion failure probability.
Note that this dictionary construction process affects neither the performance nor the privacy guarantees of our system, since it is performed by the Dictionary Provider and takes place before any queries arrive. 
The dictionary is always assumed to be known to the adversary.

To test the existence of an element $x_i$, we need only calculate its four candidate positions in $Y$, and check if any of these contain $x_i$. 
We use a Cuckoo hash table of $1.03n$ slots, storing a hash of $(\varepsilon+2)$ bits in each slot. The FPR is
$4\cdot 2^{{-(\varepsilon+2)}} = 2^{{-\varepsilon}}$.

\noindent\textbf{Query representation.} 
For each query, TA applies the four hash functions to the query to compute the $4$ possible positions of the query in $Y$, and adds these positions into the sorted list $S$.

\noindent\textbf{Carousel processing.} 
Algorithm~\ref{algo:cuckoo} shows how TA does carousel processing for the current chunk of the Cuckoo hash table. $R$ is a list of  $(\varepsilon+2)$-bit values, initialized to zeros, to store results. For each entry in $Y$, TA checks whether it is contained in $S$. If so, TA copies the byte to $R$. Otherwise, it copies it to a dummy location $dummy\_value$. It is clear that TA performs an equal number of operations for each entry in $Y$.

\noindent\textbf{Response construction.}
TA links a query back to its query presentation, and compares it with the four corresponding values in $R$.
A match with any of these indicates that the query is most probably in the dictionary $X$.

\begin{algorithm}[ht]
\caption{Membership test using \emph{Cuckoo Hash}}
\label{algo:cuckoo}
\begin{algorithmic}
    \State $Y$: Dictionary representation
    \State $S$: Query representation
    \State $R$: A list ($\varepsilon$+2)-bit empty values
    \State $dummy\_value$: a byte used to do dummy operations 
    \State $dummy\_int$: an integer used to do dummy operations 
    \State $i = 0, \quad j = 0$
    \While{$i$ is in the current chunk}
      \If {$S[j]$ equals $i$}
        \State $R[j] \gets Y[i]$
        \State $j++$ 
      \Else       
        \State $dummy\_value \gets Y[i]$
        \State $dummy\_int++ $
      \EndIf
      \State $i++$
    \EndWhile
\end{algorithmic}
\end{algorithm}

\subsection{Comparison}

Table~\ref{tbl:database-size} shows a comparison of these three representations. 
For a dictionary of $n= 2^{26}$ entries, and FPR of $2^{-10}$ the sequence of differences takes the least space.
When the number of concurrent queries $m$ is smaller than the dictionary size $N$, Cuckoo hash will be fastest (asymptotically) to process $m$ queries. 
In comparison, an ORAM-based approach requires $O(m\log N)$ time to process $m$ queries.
Asymptotically, this will be worse than the carousel approaches at about the point where  $m > N/\log N$.
We compare the performance of all methods in Section~\ref{sec:experimental_evaluation}.

\newcolumntype{J}{>{\raggedright\arraybackslash}X}
\newcolumntype{K}{>{\raggedright\arraybackslash}p{20mm}}
\newcolumntype{L}{>{\raggedright\arraybackslash}p{14mm}}

\begin{table}[t]
\caption{Comparison of dictionary representations.}
\label{tbl:database-size}
\centering
\renewcommand{\arraystretch}{1.3}
\begin{tabularx}{\columnwidth}{||>{\raggedright\arraybackslash}X||>{\raggedright\arraybackslash}p{16mm}|>{\raggedright\arraybackslash}p{13mm}|>{\raggedright\arraybackslash}p{17mm}||}
\hline
\hline
\textbf{Dictionary Representation} & \textbf{Dictionary size $(N)$} & \textbf{Size for \mbox{$\varepsilon=10$} \mbox{$n=2^{26}$}} & \textbf{Time for processing $\mathbf{m}$  queries} \\
\hline
\hline
\textbf{Sequence of Differences} & $1.02(\varepsilon + 2)n$ & 97.74 MB  & $\mathcal{O}(N\log m)$ \\ 
\hline
\textbf{Bloom Filter} & $1.44\varepsilon n$ & 115.2 MB &$\mathcal{O}(10m+N)$ \\ 
\hline
\textbf{4-ary Cuckoo Hash} & $1.03(\varepsilon+2)n$ & 98.88 MB & $\mathcal{O}(4m+N)$  \\ 
\hline
\hline
\end{tabularx}
\end{table}

\pagebreak

\section{Experimental Evaluation}
\label{sec:experimental_evaluation}

To evaluate the performance of our carousel approach, we implemented the full system (including multiple data structures) on the two most prominent hardware security architectures currently available: ARM TrustZone and Intel SGX.
In order to compare our approach with ORAM, we also implemented the essential components of a functional Path ORAM prototype on both hardware platforms.
All measurements were obtained using real hardware.

\subsection{Environment Setup}

\noindent\textbf{\KinibiTZ.}
We used a Samsung Exynos 5250 development board from Arndale with a 1.7~GHz dual-core ARM Cortex-A17 processor to implement the lookup server.\footnote{\url{http://arndaleboard.org/}}
It runs Android OS (version 4.2.1) as the host OS and Kinibi OS as the TEE OS.
Kinibi allows authorized trusted applications to execute inside the TEE.
We use the ARM GCC compiler and Kinibi-specific libraries.

Since Kinibi limits total TA private memory to a total of 1~MB, the memory available for heap and stack data structures is only about 900~KB.
This limits the number of queries that can be processed concurrently. 
Further, Kinibi on the development board only allows CA to share 1~MB of additional memory with TA.
We used this memory to transfer chunks of $Y$ as well as to submit queries and retrieve responses.
This placed an upper bound on the size of the chunks shared with TA at a given time.
CA includes the metadata (e.g. queries per chunk, chunk identifiers, and number of items from $Y$ in the  chunk) in TA invocation parameters.
To obtain timing measurements, we used the \emph{gettimeofday()} function, a Linux system call, which provides $\mu$s resolution.\footnote{\url{http://man7.org/linux/man-pages/man2/gettimeofday.2.html}}

\noindent\textbf{Intel SGX.}
We used an SGX-enabled HP~EliteDesk 800 G2 desktop PC with a 3.2~GHz Intel Core~i5~6500 CPU and 8~GB of RAM.
It runs Windows~7 (64~bit version) as the host OS, with a page size of 4~KB.
We used the Microsoft C/C++ compiler and the Intel SGX SDK for Windows.
Since we are practically unconstrained by code size, we configured the compiler to optimize execution speed (\texttt{O2}) and used the same compiler options for all experiments. 
To obtain timing measurements, we used the Windows \emph{QueryPerformanceCounter} (QPC) API, which provides high resolution (\textless1$\mu$s) time stamps suitable for time-interval measurement.\footnote{\url{https://msdn.microsoft.com/en-us/library/windows/desktop/ms644904}}

For Intel SGX, we have to account for the fact that TA does not have \emph{private memory}, and thus the adversary can observe TA's memory access pattern at page-level granularity (as discussed in Section~\ref{sec:preliminaries}).
To overcome this challenge, we designed each SGX TA such that its memory access pattern does not depend on any secret data.
A central primitive in these designs is a page-sized data container, which we refer to as an \emph{oblivious page}.
Whenever a private data structure spans more than one oblivious page, we perform the same memory access operations on all pages.
Since we assume the adversary can also measure the timing between these memory accesses, we ensure that this does not depend on and private information.
These challenges were also recently identified by Gupta et al.~\cite{Gupta2016}, who used a similar approach of ensuring constant-time operations and performing uniform memory accesses to avoid leaking information. 
We do not attempt to defend against probabilistic cache-based side-channel attacks, but we argue that these would not be feasible against our implementation since we do not perform repeated operations using any piece of secret data (e.g. as required for the attacks by Liu et al.~\cite{Liu2015b}).
Our mitigation techniques could be adapted to these types of attacks, but this would have an equal impact on the performance of all data structures in both our carousel and ORAM experiments, so the overall comparisons and conclusions will remain unchanged.

\subsection{Implementing PMT: Carousel Methods}
\label{subsec:experimentdesign:impl}

For all hash table lookups, we used the fast non-cryptographic \emph{lookup3} hash function from the set of Jenkins hash functions.\footnote{\url{http://www.burtleburtle.net/bob/c/lookup3.c}} 
For AES operations, we used the \emph{mbed~TLS} cryptography library\footnote{\url{https://tls.mbed.org/}} on \KinibiTZ{}, and the official Intel-supplied trusted cryptography library (\texttt{sgx\_tcrypto}) on SGX.
We generated a dictionary of $n=2^{26}$ items, each represented as 128~bits, drawn from a uniform random distribution.\footnote{In a real deployment, this could be a hash of a mobile application package, which is customarily used by anti-malware vendors as a (statistically) unique package identifier.}
We used the data structures described in Section~\ref{sec:dictRep} to generate $Y$.
Each user communicates with TA via a secure channel, using 128~bit AES encryption in CBC mode.

In all cases, we aimed to implement the dictionary representation data structures using an integer number of bytes so as to avoid additional bit-shift operations.
However, for the sequence of differences and 4-ary Cuckoo hash in \KinibiTZ{}, we represented items in $Y$ as 12-bit structures ($\varepsilon=10$) and operated on two items (3~bytes) at a time.
Furthermore, we optimized our implementations to make use of the largest available registers on each platform (32-bit on \KinibiTZ{} and 64-bit on Intel SGX) for read/write operations.



\noindent\textbf{Sequence of Differences.} 
Each dictionary entry was truncated to a 36-bit value ($h_i$) whilst maintaining the desired FPR ($\varepsilon=10$). 
Entries in $Y$ are 12-bit differences between two successive dictionary entries.
%
In \KinibiTZ{}, $Q$ is a linked-list ordered by chunk identifier while $S$ is maintained as a sorted array. 
Both $Q$ and $S$ are stored entirely in TA's private memory, which can accommodate a maximum of 12800 queries.
In SGX, $Q$ is stored as a sorted array spanning one or more oblivious pages.
Given the size of a query and its associated metadata (i.e. query ID and result), a single oblivious page can accommodate up to 500 queries.
If the number of concurrent queries exceeds 500, TA uses multiple oblivious pages but always performs the same number of operations on each page.
This is achieved by including a dummy query on each oblivious page.
The adversary is unable to distinguish these dummy operations from real operations since they take exactly the same amount of time and access the same oblivious page. 
Clearly, this results in many additional operations and thus has a significant impact on performance as the number of queries increases.
However, one optimization, which arises from the requirement to perform the same operations on each page, is that we can process each page independently (i.e. each page can be processed as if it were the only page present).
Although this does not negate the performance overhead, it is a significant improvement over a naive implementation.

\noindent\textbf{4-ary Cuckoo Hash.}
We use Cuckoo hash with 4 hash functions to generate $Y$. 
We represent a query as a 12-bit value, and each of the four positions as 32-bit values.
Each query representation therefore consists of a 32-bit position and a 16-bit buffer ($R$) to store the dictionary item corresponding to that position.
In \KinibiTZ{}, we maintain $S$ as a sorted linked list. The private memory can accommodate a maximum of 4500 queries.
In SGX, we again store $S$ as a sorted array spanning one or more oblivious pages.
In this scheme, we can only accommodate up to 170 queries on each oblivious page, since we must store four positions for each query.
As in the previous scheme, if the number of queries exceeds this threshold, multiple oblivious pages are used, and must all be accessed uniformly.
In addition to the previous optimization of treating these pages independently, we can further optimize by selecting hash functions that do not overlap with each other.
Fotakis et al.~\cite{DBLP:journals/mst/FotakisPSS05} used this approach to simplify the algorithm, but in our case it can also provide a significant performance advantage.
Using four non-overlapping hash functions essentially allows us to partition the dictionary representation into four regions, and consider only the query representations for one region at a time.
We therefore allocate the four query representations to four different sets of oblivious pages, thus allowing up to 680 queries per set of four pages.
When a particular region of the dictionary representation is being processed, we only operate on the pages corresponding to that region (if there are multiple such pages, the memory access must still be uniform for each of them).

\noindent\textbf{Bloom filter.}
We use Bloom filter with 10 hash functions, and thus represent each query as ten positions in $Y$.
Each query representation consists of a 32-bit position value and an 8-bit buffer ($R$) to store the byte from $Y$ which contains that position.
In \KinibiTZ{}, both $Q$ and $S$ are maintained as linked-lists in the TA's private memory. The private memory can hold a maximum of 1750 queries.
The Bloom filter approach always requires more operations than 4-ary Cuckoo hash. 
Having confirmed this on \KinibiTZ{}, we omit the repetition of this experiment on SGX.
The implementation follows the same principles as that of the Cuckoo hash scheme.



\subsection{Implementing PMT: Cuckoo-on-ORAM}
\label{subsec:experimentdesign:CoO}
Since ORAM itself is not specifically designed for PMT, we need to generate a suitable dictionary representation ($Y$) to store in the ORAM database.
We chose Cuckoo hash because it requires the fewest memory accesses. 
By comparison, each Bloom filter query requires 10 different accesses, and each binary search in the sequence of differences representation accesses at least 26 positions. 
TA is the ORAM \emph{processor} whilst CA stores the encrypted shuffled database. 
When TA receives a query, it maps the query to four cuckoo positions in $Y$. 
It then access these four positions following an ORAM protocol to complete the PMT. 
Since ORAM was designed to hide the access patterns, the adversary \Adv{} learns no information about which positions have been accessed.
We chose Path-ORAM as baseline for comparison because of its simplicity and because the Goldreich-Ostrovsky lower bound of $\mathcal{O}(m\log~n)$ amortized lookups for $m$ queries applies to all ORAM variants mentioned in Section~\ref{sub:sec:oram}, e.g. Ring-ORAM only has a 1.5x speedup over Path-ORAM in the secure-processor setting. Moreover, advanced parallel or asynchronous ORAM schemes require parallel computation, and are difficult to implement without leaking information, e.g. TaORAM requires additional temporary data storage in TA, which in SGX must be made oblivious. A summary of our chosen parameters is shown in Table~\ref{tbl:oram-parameters}.

\begin{table}[ht]
\renewcommand{\arraystretch}{1.3}
\caption{Path ORAM parametrization.}
\label{tbl:oram-parameters}
\centering
\begin{tabular}{||l|l|l|l||}
\hline \hline

\textbf{Block size}
&

\textbf{Node size} 
&
\textbf{Tree size} 
& 
\textbf{Tree height}
\\ \hline \hline

4~KB 
& 
4 blocks    
& 
6329 nodes  
&
13 levels \\
\hline \hline


\end{tabular}
\end{table}

We set the block size to 4~KB and each node of the tree contains 4 blocks.
Our 98.88~MB dictionary (Table~\ref{tbl:database-size}) therefore required 6329 nodes, which results in a tree of height 13.
It required a 6~KB position map which can easily be stored in TA's private memory.


Although the Path ORAM algorithm is relatively simple, implementing it in full has been found to be quite complicated~\cite{Bindschaedler2015}, and is not required for this comparison.
Instead, we prototyped the main operations and in all cases chose options that favor the ORAM implementation.
This partial prototype therefore represents a generous upper bound on the performance of any full implementation.

\noindent\textbf{\KinibiTZ{}.}
In \KinibiTZ{}, we avoided maintaining the \emph{stash} required by Path ORAM. 
Instead while storing a path back, the nodes were re-encrypted and shuffled along the path and the position map updated accordingly.
Again, this simplification favors ORAM in the comparison since maintaining a stash would increase the number of operations performed per query.

\noindent\textbf{Intel SGX.}
Since Path ORAM assumes some amount of private memory, which is not available in SGX, we had to take additional steps to ensure that no information is leaked through the enclave's memory access pattern.
As with previous schemes, we used the concept of an oblivious page.
All private data structures are stored on oblivious pages, and whenever a data structure's size exceeds one page, we ensure that the same sequence of operations is performed on each page (e.g. by reading and writing dummy values).

Specifically, with the above parameters, the Path ORAM position map spanned four oblivious pages, thus requiring four reads/writes for every read/write to the position map.
Each node in the stash also takes up four pages.
Reading a node into the stash does not require specific privacy protection (e.g. \Adv{} may learn the location of a specific node in the stash without compromising privacy), and thus no additional operations are required.
However, whenever a node is evicted from the stash, \Adv{} must not be able to identify the evicted node.
To achieve this, we allocate a stash output buffer, equal to the size of one node, within the enclave's secure memory. 
We then iterate over all nodes in the stash, copying the intended node into the output buffer and performing a constant-time dummy write to the output buffer for every other block.
Since the stash output buffer is still in the enclave's secure memory, \Adv{} cannot determine which node has been placed in this buffer.
The contents of the output buffer are then encrypted and evicted as usual.\footnote{The issue of preventing side-channels from leaking information about the ORAM queries is similar to the issue of asynchronicity in ORAM queries discussed in~\cite{SZE16}. 
We took a conservative approach of preventing such side-channels.}

We use the same optimization for Cuckoo hash as described earlier: the four hash functions are selected to have non-overlapping outputs.
As above, this allows us to partition the dictionary representation into four different regions.
In the case of ORAM, we construct four separate ORAM trees, such that each holds the values for a single region.
This optimization improves performance in the Path ORAM case by reducing the size of each tree, and hence the path length and size of the position map.
With this optimization, each tree's position map fits onto two oblivious pages.


\subsection{Performance evaluation}
\label{subsec:experimentdesign:eval}

\begin{figure*}[t]
\centering

%
%
%



\begin{tikzpicture}
	\begin{axis}[
		width=16cm,
		height=5cm,
		xlabel={Number of queries},
		ylabel={Processing time (seconds)},
		scaled ticks=false,
		y tick label style={
			/pgf/number	format/fixed
		},
		legend style={
         	overlay,
         	at={(0.8,0.75)
         },
		anchor=center},
		ytick={1,5,10,15,20},
		minor ytick={1,2,...,20},
		xmin=-0,
		xmax=4500,
		ymin=-0,
  		ymax=20,grid=both,
  		grid style={line width=.1pt, draw=gray!10},
    	major grid style={line width=.2pt,draw=gray!50}]

\addplot+[error bars/.cd,y dir=both,y explicit][blue,  mark=triangle,mark options={blue, scale=1}]coordinates {

	(1, .43) +- (0, .13)  
	(10, .47)+-(0, .17) 
	(100, .49)+-(0, .16)
	(500, .81) +-(0, .19) 
	(1000, 1.80) +-(0, .33)
	(1500, 3.81) +- (0, .45) 
	(1750, 4.72) +-(0, .49)
};
\addlegendentry{Bloom-Filter-on-a-Carousel}

\addplot+[error bars/.cd,y dir=both,y explicit][forest, mark=square* ,mark options={forest, scale=0.5}]coordinates {

	(1, .73)  +-(0, .16) 
	(500, .75)  +-(0, .12) 
	(1000, 1.13) +-(0, .42)
	(1500, 1.49) +-(0, .53)
	(2000, 1.83) +-(0, .59)
	(2500, 2.19) +-(0, .43)
	(3000, 2.57) +-(0, .52)
	(3500, 2.92) +-(0, .53)
	(4000, 3.27) +-(0, .60)
};
\addlegendentry{Cuckoo-on-a-Carousel}

\addplot+[error bars/.cd,y dir=both,y explicit][blues4,  mark=x, mark options={solid, scale=0.7}]coordinates {
(10, 4.97) +-(0, 1.1) 
(500, 8.00) +-(0, 1.3)
(1000, 8.76) +-(0, 1.12)
(1500, 9.54) +-(0, .65)
(2000, 9.95) +-(0, 1.2)
(3000, 11.84) +-(0, 1.73)
};
\addlegendentry{Differences-on-a-Carousel}

\addplot+[error bars/.cd,y dir=both,y explicit][carmine,  mark=*, mark options={solid, scale=0.6}]coordinates {
(1, .009)	 +-(0, .01)
(10, .09) +-(0, .08) 
(100, .89) +-(0, .26)
(1000, 9.21) +-(0, .53)
(1500, 13.81) +-(0, .59)
(2000, 18.63) +-(0, .62)
};
\addlegendentry{Cuckoo-on-ORAM}

	\end{axis}
\end{tikzpicture}
\caption{\KinibiTZ: Total processing time for a batch of queries (average and variance over 1000 runs).}
\label{fig:kinibi_carousel}
\end{figure*}
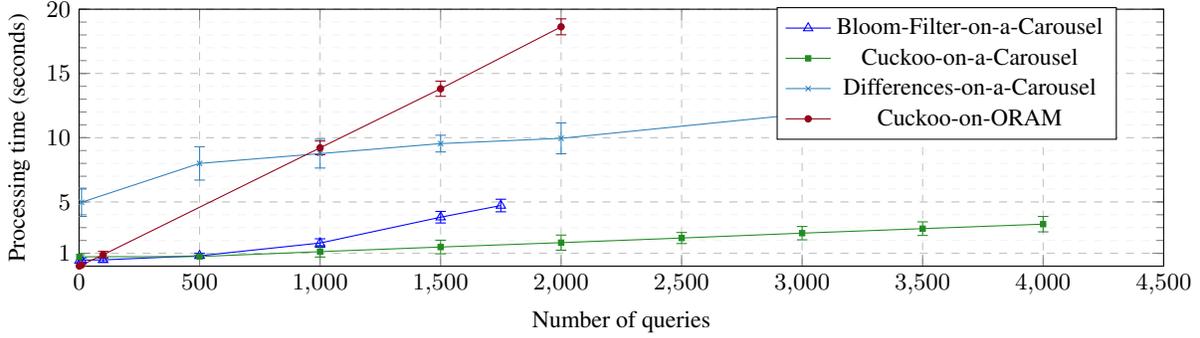

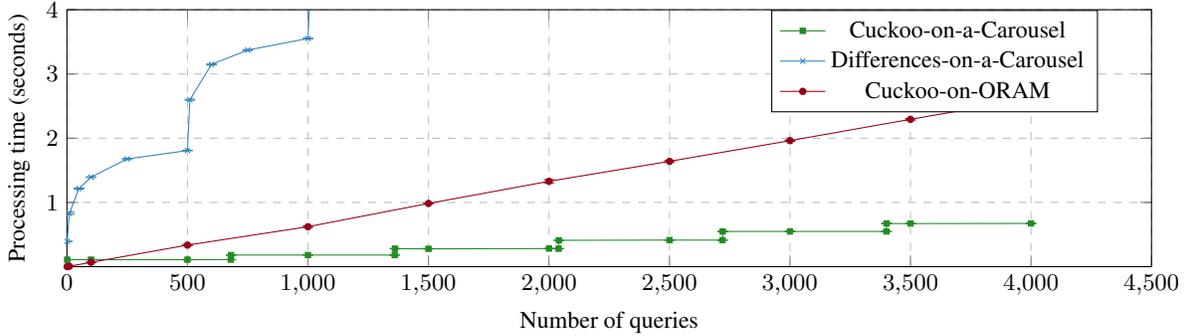
\begin{figure*}[t]
\centering

%
%
%




\begin{tikzpicture}
	\begin{axis}[
		width=16cm,
		height=5cm,
		xlabel={Number of queries},
		ylabel={Processing time (seconds)},
		scaled ticks=false,
		y tick label style={/pgf/number	format/fixed},
		legend style={overlay,at={(0.8,.80)},anchor=center},
		ytick={1,2,3,4,5,6},
		minor ytick={1,2,...,12},
		xmin=-0,
		xmax=4500,
		ymin=-0,
  		ymax=4, 
  		grid=both,
  		grid style={line width=.1pt, draw=gray!10},
    	major grid style={line width=.2pt,draw=gray!50}]

\addplot+[error bars/.cd,y dir=both,y explicit][forest, mark=square* ,mark options={forest, scale=0.5}]coordinates {
	(1,0.109) +- (0,0.002)  
	(100, 0.109) +-(0,0.002)
	(500, 0.109) +- (0,0.003)
	(680, 0.110) +-(0,0.001) 
	(681, 0.181) +-(0,0.004) 
	(1000, 0.181) +-(0,0.003)
	(1360, 0.181) +-(0,0.003)
	(1361, 0.280) +-(0,0.004)
	(1500, 0.278) +-(0,0.004) 
	(2000, 0.282) +-(0,0.003)
	(2040, 0.280) +-(0,0.003)
	(2041, 0.410) +-(0,0.003)
	(2500, 0.414) +-(0,0.004)
	(2720, 0.413) +-(0,0.005)
	(2721, 0.548) +-(0,0.005)
	(3000, 0.548) +-(0,0.004)
	(3400, 0.548) +-(0,0.003)
	(3401, 0.672) +-(0,0.006)
	(3500, 0.671) +-(0,0.004)
	(4000, 0.673) +-(0,0.005)
};
\addlegendentry{Cuckoo-on-a-Carousel}

\addplot+[error bars/.cd,y dir=both,y explicit][blues4,  mark=x, mark options={solid, scale=0.7}]coordinates {
	(1,0.3932) +- (0,0.003) 
	(10, 0.8310) +- (0,0.016) 
	(50, 1.2171) +- (0,0.008) 
	(100, 1.3938) +-(0,0.005)
	(250, 1.6771) +- (0,0.009) 
	(500, 1.8082) +-(0,0.004)
	(510, 2.5990) +- (0,0.007) 
	(600, 3.1486) +- (0,0.007) 
	(750, 3.3724) +- (0,0.007)
	(1000, 3.5520) +-(0,0.006) 
	(1010, 4.3452) +- (0,0.013) 
	(1100, 4.8939) +- (0,0.007) 
	(1250, 5.1179) +-(0,0.004)
	(1500, 5.2959) +-(0,0.010) 
	(1510, 6.0849) +- (0,0.017) 
	(1600, 6.6530) +- (0,0.016) 
	(1750, 6.8629) +-(0,0.009)
	(2000, 7.0377) +-(0,0.005) 
	(2010, 7.8516) +- (0,0.012) 
	(2100, 8.4064) +- (0,0.010) 
	(2250, 8.6184) +-(0,0.007)
	(2500, 8.8000) +-(0,0.006)
	(2510, 9.6080) +- (0,0.011) 
	(2600, 10.1649) +- (0,0.009) 
	(2750, 10.3845) +-(0,0.008)
};
\addlegendentry{Differences-on-a-Carousel}

\addplot+[error bars/.cd,y dir=both,y explicit][carmine,  mark=*, mark options={solid, scale=0.6}]coordinates {
(   1, 0.001) +-(0, 0.000)
(  10, 0.007) +-(0, 0.001) 
( 100, 0.067) +-(0, 0.001)
( 500, 0.336) +-(0, 0.010)
(1000, 0.622) +-(0, 0.011)
(1500, 0.984) +-(0, 0.005)
(2000, 1.329) +-(0, 0.025)
(2500, 1.640) +-(0, 0.011)
(3000, 1.962) +-(0, 0.009)
(3500, 2.294) +-(0, 0.011)
(4000, 2.618) +-(0, 0.008)
};
\addlegendentry{Cuckoo-on-ORAM}


	\end{axis}
\end{tikzpicture}

\caption{Intel SGX: Total processing time for a batch of queries (average and variance over 1000 runs).}
\label{fig:sgx_carousel}
\end{figure*}

\noindent\textbf{Batch Performance.}
Figures~\ref{fig:kinibi_carousel} and~\ref{fig:sgx_carousel} show the total processing time for a single batch of queries using different carousal schemes. 
Queries were sent in a batch at the beginning of the each carousel cycle.
To achieve the desired FPR, we used dictionary representations with $\varepsilon=10$ on \KinibiTZ{}, and $\varepsilon=14$ on Intel SGX (since it was more efficient to operate on byte-aligned data structures on the Intel SGX platform).

Each point in the figures represents the average time for processing the batch over 1000 repetitions. 
The figures show that processing time increases with query load for all three carousel schemes.
On both platforms, Difference-on-a-carousel has longer processing time, because of having to do a binary search on $S$ for every item in $Y$.
On Intel SGX, the non-linear step-like behavior is caused by the use of multiple oblivious pages.
Since the same number of operations must be performed on each page (to preserve privacy), each additional page causes a step increase in processing time.
The width of each step corresponds to the number of queries that can be accommodated per page.
In the Difference-on-a-carousel scheme, the steps take a logarithmic shape due to the binary search on the final (under-utilized) page, which eventually reaches full capacity. 

On \KinibiTZ{}, under small query load (less than 500 queries), the batch processing time for Bloom-filter-on-a-carousel is faster than other carousel approaches, however, processing time increases rapidly as the number of queries grows (beyond 1000 queries).
The hardware was unable to support larger query batch sizes.

Cuckoo-on-a-carousel (CoaC) can process more queries with less overhead than the other methods.
Again, the non-linear performance characteristics in SGX are due to the use of multiple oblivious pages.
Since this algorithm requires only pointer-based operations (i.e. no binary search), each step adds a constant number of additional operations, resulting in flat step increase.

In contrast to CoaC, Cuckoo-on-ORAM (CoO) provides a very fast response latency (9~ms) for a single query.
However, queries are processed sequentially. 
For example, when 2,000 queries arrive at once, the latency of the final response will be 18 seconds on \KinibiTZ{}, which is beyond the acceptable tolerance of a malware checking service.
By comparison, on \KinibiTZ{}, CoaC takes only 1.83 seconds to process 2,000 queries.
Results for Intel SGX show a similar pattern with significantly lower latencies (e.g. 0.282 seconds to process 2,000 queries).
Carousel schemes therefore achieve lower query response latencies when handling batches of queries.

%
%

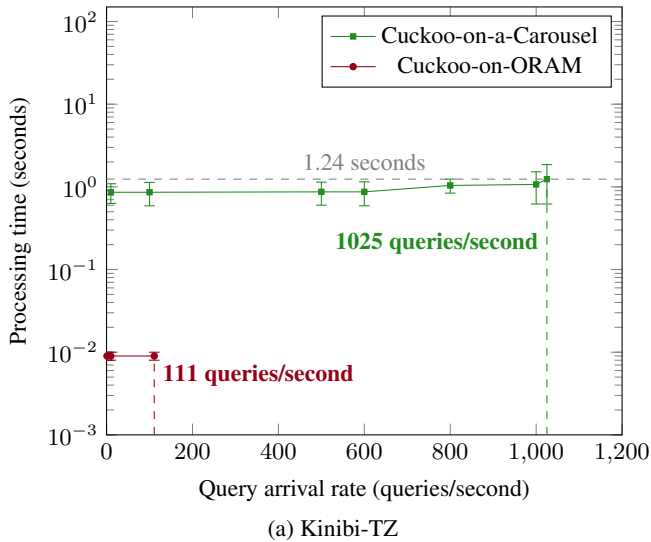
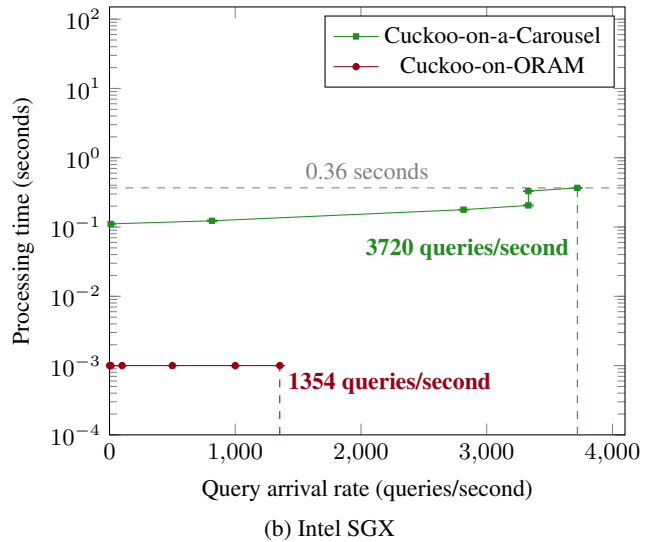
\begin{figure*}[t]
\subfloat[\KinibiTZ]{

%
%
%




\begin{tikzpicture}
	\begin{axis}[
		xlabel={Query arrival rate (queries/second)},
		ylabel={Processing time (seconds)},
		scaled ticks=false,
		y tick label style={/pgf/number	format/fixed},
		ymode=log,
		xmin=-0,
		xmax=1200,
		ymin=0.001,
		ymax=150]

\addplot+[error bars/.cd,y dir=both,y explicit][forest, mark=square* ,mark options={forest, scale=0.5}]coordinates {
(   10, 0.86)   +-(0, 0.23)
(  100, 0.86) 	+-(0, 0.27)
(  500, 0.87)	+-(0, 0.27)
(  600, 0.87)   +-(0, 0.28)
(  800, 1.04)	+-(0, 0.20)
( 1000, 1.07)	+-(0, 0.45)
( 1025, 1.24)   +-(0, 0.62)
};\label{coac}
\addlegendentry{Cuckoo-on-a-Carousel}

\addplot+[error bars/.cd,y dir=both,y explicit][carmine, mark=*, mark options={solid, scale=0.6}]coordinates {
(   1, 0.009) +-(0, 0.001)
(  10, 0.009) +-(0, 0.001) 
( 111, 0.009) +-(0, 0.001)
};\label{coo}
\addlegendentry{Cuckoo-on-ORAM}

\addplot[mark=none, forest, dashed] coordinates {(1025, \pgfkeysvalueof{/pgfplots/ymin}) (1025, 1.24)} node[left,pos=0.75] {\textbf{1025 queries/second}};
\addplot[mark=none, carmine, dashed] coordinates {(111, \pgfkeysvalueof{/pgfplots/ymin}) (111, 0.009)} node[right,pos=0.75] {\textbf{111 queries/second}};
\addplot[mark=none, gray, dashed] coordinates {(\pgfkeysvalueof{/pgfplots/xmin},1.24) (\pgfkeysvalueof{/pgfplots/xmax},1.24)} node[above,pos=0.5]{1.24 seconds};

\end{axis}




\end{tikzpicture}

\label{kinibiQR_arrival_rates}}
\hfill
\subfloat[Intel SGX]{

%
%
%




\begin{tikzpicture}
	\begin{axis}[
		xlabel={Query arrival rate (queries/second)},
		ylabel={Processing time (seconds)},
		scaled ticks=false,
		y tick label style={/pgf/number	format/fixed},
		xmin=0,
		xmax=4100,
		ymode=log,
		ymin=0.0001,
  		ymax=150,]

\addplot+[error bars/.cd,y dir=both,y explicit][forest, mark=square* ,mark options={forest, scale=0.5}]coordinates {
	(9.03,    0.111) +- (0,0.003)  
	(814.66,  0.123) +-(0,0.003)
	(2815.73, 0.178) +- (0,0.002)
	(3331.32, 0.205) +-(0,0.004) 
	(3331.33, 0.3293) +-(0,0.004)
	(3720.95, 0.367) +-(0,0.005)
};
\addlegendentry{Cuckoo-on-a-Carousel}

\addplot+[error bars/.cd,y dir=both,y explicit][carmine,  mark=*, mark options={solid, scale=0.6}] coordinates {
(   1, 0.001) +-(0, 0.000)
(  10, 0.001) +-(0, 0.000) 
( 100, 0.001) +-(0, 0.000)
( 500, 0.001) +-(0, 0.000)
(1000, 0.001) +-(0, 0.000)
(1354.28, 0.001) +-(0, 0.000)
};
\addlegendentry{Cuckoo-on-ORAM}

%
%
%
%

\addplot[mark=none, carmine, dashed] coordinates {(1354.28,\pgfkeysvalueof{/pgfplots/ymin}) (1354.28,0.001)} node[right,pos=0.75] {\textbf{1354 queries/second}};
\addplot[mark=none, forest, dashed] coordinates {(3720.95,\pgfkeysvalueof{/pgfplots/ymin}) (3720.95,0.367)} node[left,pos=0.75] {\textbf{3720 queries/second}};
\addplot[mark=none, gray, dashed] coordinates {(\pgfkeysvalueof{/pgfplots/xmin},0.367) (\pgfkeysvalueof{/pgfplots/xmax},0.367)} node[above,pos=0.5] {0.36 seconds};


	\end{axis}
\end{tikzpicture}


\label{sgxQR_arrival_rates}}
\caption{Steady-state processing time for uniform query arrival rates (average and variance over 1000 runs). Vertical lines indicate breakdown points.}
\label{fig:uniformQR}
\end{figure*}

\noindent\textbf{Steady-state Performance.}
In addition to measuring batch query processing, we also compare the steady-state performance of CoaC and CoO, assuming a constant query arrival rate.
Again we are primarily concerned with the average query response latency.
On \KinibiTZ{}, CoO provides responses with a latency of 9~ms if the arrival rate is below 111 queries/second.
On Intel SGX, this latency decreases to 1~ms latency for arrival rates below 1000 queries/second.
Figure~\ref{kinibiQR_arrival_rates} and Figure~\ref{sgxQR_arrival_rates} show the steady state performance of \KinibiTZ{} and Intel SGX for different query arrival rates (averaged over 1000 repetitions).
%
%
%
In order to identify the breakdown point where CoaC can no longer guarantee a bounded query response latency, we simulated the steady-state operation of CoaC with different query rates and calculated the average number of concurrent queries in TA (i.e. the \emph{occupancy}) during each carousel cycle.
%
On \KinibiTZ{}, we identify a query rate as sustainable when the average query occupancy remains stable at a level below the maximum number of concurrent queries the TA can handle (e.g. 4500 queries).
We noticed that the carousel cycle time fluctuates due to OS scheduling, which occasionally causes the occupancy to reach the maximum capacity. 
Although occasional spikes can be tolerated, we consider the breakdown point to be the arrival rate at which the average occupancy consistently reaches this maximum capacity. 
\ifsubmission
For example in \KinibiTZ{}, we found that for query arrival rates above 1030 queries/second, the CoaC query response latency cannot be sustained.
In contrast, with 1025 queries/second, the operating characteristics of CoaC are stable, and we therefore conclude that the breakdown point is between 1025 and 1030 queries/second. 
\else
Figure~\ref{occupancy_rate1025} and Figure~\ref{occupancy_rate1030} shows the evolution of query occupancy in the \KinibiTZ{} beyond 500 carousel cycle. 
At 1030 queries/second (Figure~\ref{occupancy_rate1030}), the CoaC query response latency cannot be sustained. In contrast, with 1025 queries/second the linear regression on (Figure~\ref{occupancy_rate1030}) occupancy suggests that CoaC will provide a sustainable response latency, and we therefore conclude that the breakdown point is between 1025 and 1030 queries/second.
\fi

For Intel SGX, we noted that there is virtually no variability in the batch performance results (i.e. the results do not change much over multiple runs), and leveraged this to ascertain the steady-state breakdown point.
We set the occupancy of TA to a fixed value and measured the time taken to process one carousel's worth of chunks.
Dividing this fixed occupancy by the average carousel time gives the maximum query arrival rate sustainable at that occupancy level.
Repeating this for multiple occupancy values yields the curve in Figure~\ref{sgxQR_arrival_rates}.
For Intel SGX, this is the best method for determining the breakdown point because of the non-linear behavior caused by the oblivious pages.
Although each additional page allows more queries to be processed in a single carousel cycle, it also adds a performance penalty, which increases the carousel cycle time.
Therefore, the maximum sustainable rate can only be achieved by fully utilizing every page.
In practice, the system would employ a optimization algorithm to select the optimal number of pages for each situation.
The steady-state query rates shown in Figure~\ref{sgxQR_arrival_rates} would be the input parameters for this optimization algorithm.

\ifsubmission
\else

\begin{figure*}[p]
\centering
\centerline{\includegraphics[ width=1\linewidth]{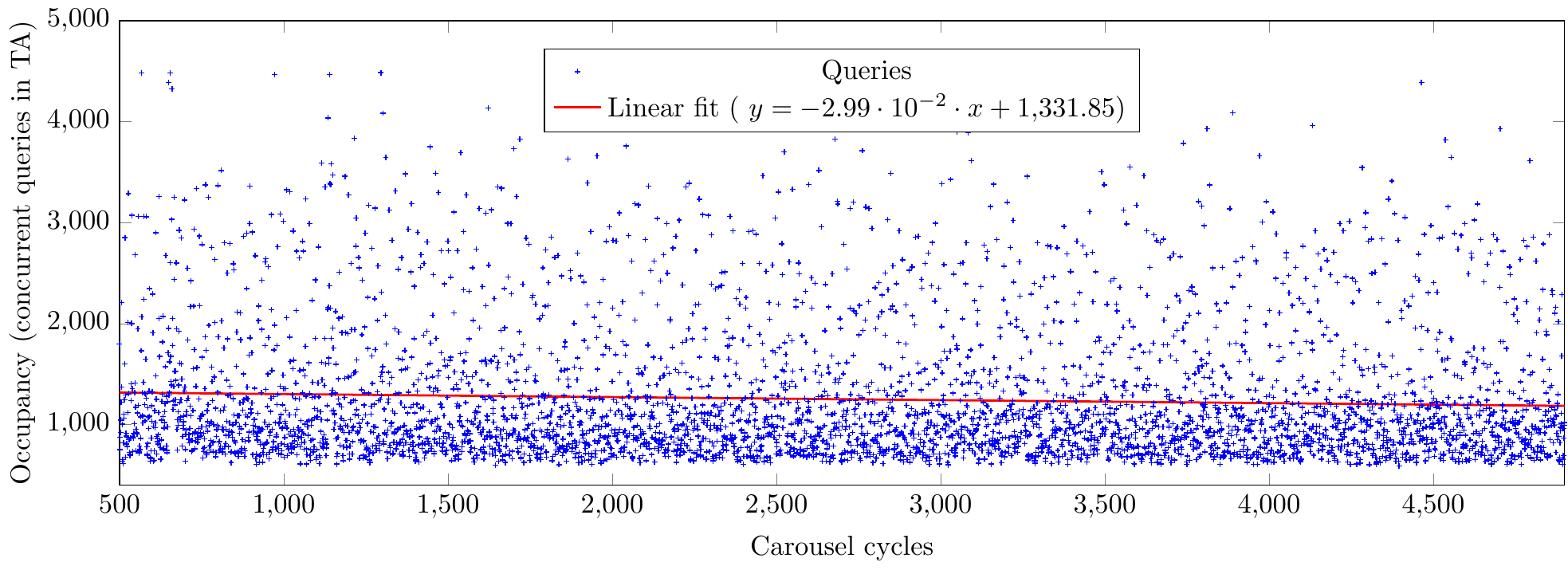}}
\caption{\KinibiTZ{}: Evolution of occupancy at 1025 queries/second}
\label{occupancy_rate1025}
\end{figure*}

\begin{figure*}[p]
\centering
\centerline{\includegraphics[, width=1\linewidth]{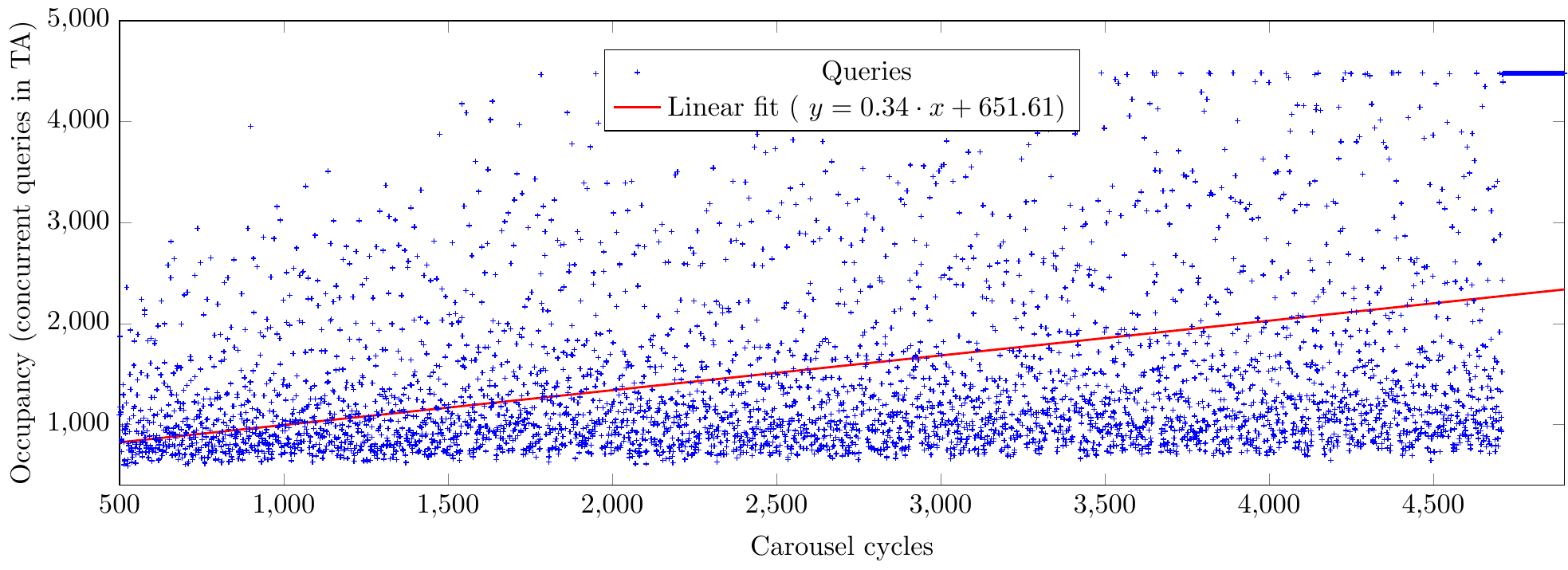}}
\caption{\KinibiTZ{}: Evolution of occupancy at 1030 queries/second}
\label{occupancy_rate1030}
\end{figure*}

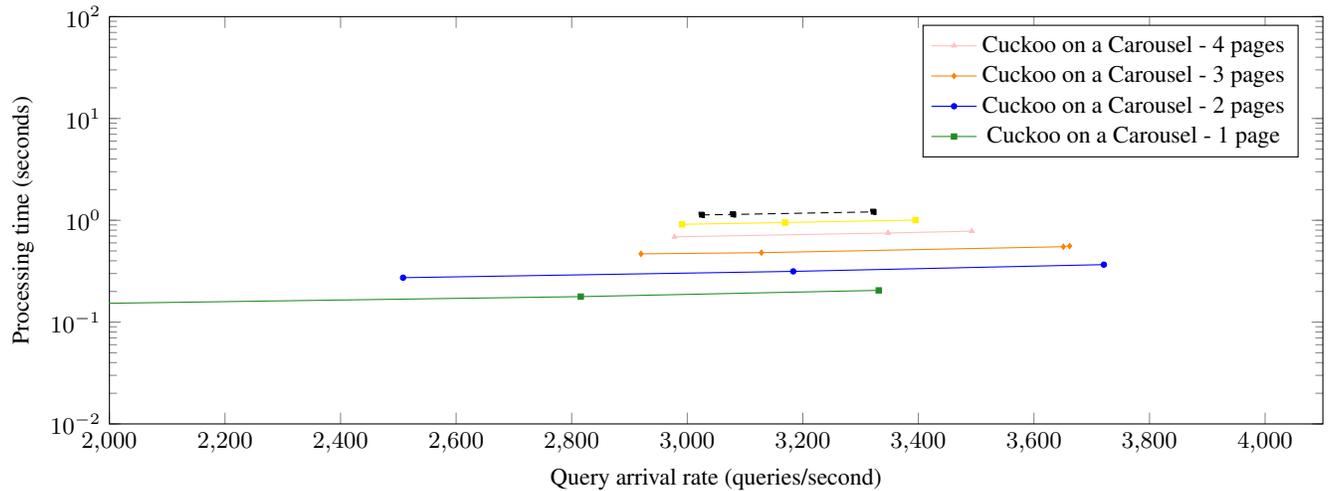
\begin{figure*}[p]
\centering
\centerline{

%
%
%




\begin{tikzpicture}
	\begin{axis}[
		width=1\linewidth,
		height=7cm,
		xlabel={Query arrival rate (queries/second)},
		ylabel={Processing time (seconds)},
		scaled ticks=false,
		y tick label style={/pgf/number	format/fixed},
		xmin=2000,
		xmax=4100,
		ymode=log,
		ymin=0.01,
  		ymax=100,]



\addplot+[pink, mark=triangle* ,mark options={pink, scale=0.5}]coordinates {
	(2978.03, 0.688) +-(0,0.003)
	(3347.39, 0.750) +-(0,0.004)
	(3492.45, 0.782) +-(0,0.005)
};
\addlegendentry{Cuckoo on a Carousel - 4 pages}

\addplot+[orange, mark=diamond* ,mark options={orange, scale=0.5}]coordinates {
	(2919.89, 0.468) +-(0,0.004)
	(3128.40, 0.481) +-(0,0.004) 
	(3650.97, 0.550) +-(0,0.003)
	(3661.81, 0.559) +-(0,0.003)
};
\addlegendentry{Cuckoo on a Carousel - 3 pages}

\addplot+[blue, mark=otimes* ,mark options={blue, scale=0.5}]coordinates {
	(2508.18, 0.273) +-(0,0.004) 
	(3183.48, 0.315) +-(0,0.003)
	(3720.95, 0.367) +-(0,0.003)
};
\addlegendentry{Cuckoo on a Carousel - 2 pages}

\addplot+[forest, mark=square* ,mark options={forest, scale=0.5}]coordinates {
	(9.03,    0.111) +- (0,0.002)  
	(814.66,  0.123) +-(0,0.002)
	(2815.73, 0.178) +- (0,0.003)
	(3331.32, 0.205) +-(0,0.001) 
};
\addlegendentry{Cuckoo on a Carousel - 1 page}

\addplot+[yellow, mark=square* ,mark options={yellow, scale=0.5}]coordinates {
	(2990.91, 0.913) +-(0,0.005)
	(3169.27, 0.950) +-(0,0.004)
	(3394.95, 1.005) +-(0,0.003)
};

\addplot+[black, mark=square* ,mark options={black, scale=0.5}]coordinates {
	(3024.95, 1.129) +-(0,0.006)
	(3079.20, 1.141) +-(0,0.004)
	(3322.11, 1.209) +-(0,0.005)
};



	\end{axis}
\end{tikzpicture}


}
\caption{Intel SGX: Steady-state response latency for uniform query arrival rates (average and variance over 1000 runs).}
\label{sgxQR_arrival_rates_2}
\end{figure*}

For Intel SGX, Figure~\ref{sgxQR_arrival_rates_2} shows part of the same graph as Figure~\ref{sgxQR_arrival_rates} with an additional set of data points.
Each curve corresponds to a different number of oblivious pages.
Note that the curves for multiple pages are not defined for lower query rates (i.e. towards the left boundary of the figure), since the algorithm will never use more pages than necessary. 
This figure shows that the maximum query rate is achieved when all the pages are fully utilized, but shows that increasing the number of pages beyond a certain point does not increase the overall maximum query rate.
When using more than two oblivious pages, the impact of the performance penalty for adding another page exceeds the benefit provided by that page, and ultimately results in a lower maximum rate.
The same trend continues for larger numbers of oblivious pages, suggesting that two pages is the optimal situation.

\fi

\section{Variations and Extensions}
\label{sec:extensions}

%

\noindent\textbf{Query scalability.} 
Query arrival rates that exceed the breakdown point can be supported by adding new hardware so that multiple TAs can run in parallel. The same dictionary representation can be replicated for each TA. 
Without loss of privacy, any incoming query can be routed to any TA (e.g. using any type of load balancing scheme) since each TA has its own dictionary representation.

\noindent\textbf{Dictionary scalability.} 
Our carousel approach is specifically designed around the parameters for the malware checking use case, including generous safety margins (e.g. a dictionary size of $2^{26}$ entries). For larger dictionary sizes, the dictionary can be split into multiple subsets, each handled by a separate TA running on its own core or processor.
To ensure query privacy, an adversary must not be able to identify which TA receives a given query. This requires a central dispatcher TA that multiplexes incoming requests to the worker TAs. 
Additional decoy traffic may be needed to thwart the adversary from gaining information via traffic analysis.

\noindent\textbf{Compact representation vs. complexity of processing.} 
More compact dictionary representations may lead to shorter carousel cycle times, but this may be offset by
the complexity of processing the representation. Conditional clauses (\texttt{if}) in the carousel processing
logic are particularly expensive. For example, we initially implemented the sequence of differences approach
using Huffman encoding to represent the differences. This resulted in each difference represented by
$\varepsilon+1.35$ bits on average, which is a significant reduction in dictionary size. In particular, as
Huffman encoding is prefix-free, there was no need to add dummy entries (as explained in
Section~\ref{sec:dictRep}). However, the decoding process required processing variable-size suffixes, which
resulted in an overall increase in the carousel cycling time.


\noindent\textbf{Implementation optimization.}
By default, items in the dictionary representations are not necessarily aligned on byte boundaries (e.g. in the sequence of differences and Cuckoo hash methods, our desired FPR results in dictionary representations with 12-bit item length).
Extracting such an item from a bit string requires multiple shift and add operations compared to byte-aligned representations.
However, in \KinibiTZ{} we still use 12-bit representations since we can represent two items with exactly 3 bytes.
Similarly, we reduced the number of read operations by designing our algorithms to read data at the maximum register size of each platform.



\noindent\textbf{Adversary capabilities.} We assumed that the adversary can observe the full memory access pattern for non-private memory (e.g. the CA's memory, from which the dictionary representation is accessed). This provides the strongest privacy guarantee on all hardware platforms. However, if certain platforms do not allow the adversary to make such detailed observations, our approach could be further optimized for these platforms without impacting privacy.

\section{Analysis}
\label{sec:analysis}

\noindent\textbf{Privacy Analysis.}
TA is implemented to behave essentially as a trusted third party. 
Namely: (1) The communication channels between users and TA are encrypted and authenticated. 
(2) Remote attestation guarantees to all parties that TA runs the required program. 
(3) The TEE isolates TA's computation from the rest of the system.  
(4) The algorithms that are used (Algorithms~\ref{algo:difference}, \ref{algo:bf} and \ref{algo:cuckoo}) were designed and carefully implemented to prevent side-channels.
In general, guaranteeing that different code paths take equal processing cannot be fully achieved at source code level. 
In Appendix~\ref{app:code_optimization}, we discuss how we ensured equal processing time at instruction level.
The access patterns and the entire behavior of TA, when viewed externally, are indistinguishable for different query sets of the same size.
In all algorithms, TA accesses every dictionary entry within the dictionary and performs an equal number of operations per entry, regardless of whether a match is found. 
An adversary who measures, for example, the time taken to process a given chunk, will always get the same measurement, since this time depends on the number of queries but not on the contents of the queries.
Therefore, Requirement~\ref{R1} is satisfied.



\noindent\textbf{Performance Analysis.}
Figure~\ref{fig:kinibi_carousel} and~\ref{fig:sgx_carousel} show that the carousel time for 1000 simultaneous queries is within about a second for both \KinibiTZ{} and SGX, satisfying Requirement~\ref{R2} (latency).
When the number of simultaneous queries in the TA increase to 4000, the response latency is still reasonable (4 seconds for \KinibiTZ{} and 2 seconds for SGX).
Figures~\ref{kinibiQR_arrival_rates} and~\ref{sgxQR_arrival_rates} show that the carousel approach can sustain a high query arrival rate (1025 queries/second for \KinibiTZ{} and 3720 queries/second for SGX) without breakdown.  
Use of multiple TEEs can support more queries or a larger dictionary, satisfying Requirement~\ref{R3} (scalability). 
Finally, none of our schemes introduces any false negatives, and the false positive rate is within the $2^{-10}$ limit identified (Requirement~\ref{R4}).

\section{Related Work}
\label{sec:related_work}


\emph{Private Information Retrieval} (PIR) is a cryptographic protocol that allows a user to retrieve an item from a known position in a server's database without the server learning which item was accessed.
The first single-server scheme was introduced by Kushilevitz and Ostrovsky~\cite{Kushilevitz1997}.
%
%
It is not reasonable to assume that users know the indices of desired items. This motivates \emph{Private Keyword Search} (PKS).
In PKS, the server holds a database of pairs $\{(x_1,p_1), \ldots, (x_n,p_n)\}$, where $x_i$ is a keyword and $p_i$ is a payload.
A query is a searchword $x$ instead of an index.
After the protocol, the user gets the result $p_i$ if there is a value $i$ for which $x_i = x$ or otherwise receives a special symbol $\perp$.
PKS can be constructed based on PIR, oblivious polynomial evaluation~\cite{Freedman2005}, or multiparty computation~\cite{Pappas2014}.
Private membership test can be viewed as a simplified version of PKS, where the user does not require the actual payload.
The main limitation of the current PIR/PKS solutions is their efficiency, in terms of both computation and communication. 

In addition to the purely cryptographic solutions, another option is to use trusted hardware combined with cryptography to solve the PIR/PKS problems.
For example, \cite{Iliev2005} can achieve PIR with constant computation and communication, but have to periodically re-shuffle the dataset.
Backes et al.~\cite{Backes2012} combine trusted hardware with ORAM to ensure access privacy in online behavioral advertising.
However, this approach has two drawbacks compared with our solution: it requires all elements in the database to be encrypted thus some subset must be decrypted to answer each query, and it is hard to achieve batched query processing, thus limiting scalability. 

Another approach for implementing PMT is to have the server offload some data to the user (retaining the same order) in the offline phase.
This allows constant communication and computation for each query in the online phase~\cite{Nojima2009, Meskanen2015}. 
However, the drawback is that it prevents the dataset from being updated frequently, which is a critical requirement for a malware checking use case.

\section{Conclusion and Future Work}
\label{sec:conclusion}

Motivated by the problem of privacy-preserving cloud-based malware checking, we introduced a new carousel approach for private membership test. 
We evaluated several data structures for representing the dictionary and described how to adapt them to the carousel design pattern. We implemented these on both ARM TrustZone and Intel SGX and found that Cuckoo hash provides the lowest query response latency. 
We compared our carousel approach with ORAM, and found that the former can sustain significantly higher query arrival rates. 
Future work will investigate other data structures for representing the dictionary, compare newer ORAM schemes, and explore new ways of using trusted hardware to enhance these schemes.

\section*{Acknowledgements}
\label{sec:ack}
This work was supported by the \emph{Cloud Security Services (CloSer)} project (3881/31/2016), funded by TEKES -- the Finnish Funding Agency for Innovation, and by a grant from the Israel Science Foundation, and the BIU Center for Research in Applied Cryptography and Cyber Security in conjunction with the Israel National Cyber Bureau in the Prime Minister's Office.

\bibliographystyle{acm}
\bibliography{pmt-th}  

\appendix

\section{Additional Benchmarks}
\label{app:additional_benchmarks}

\begin{table}[ht]
\renewcommand{\arraystretch}{1.3}
\caption{Average carousel cycle time of a 116~MB dictionary representation under different access patterns.}
\label{tbl:statistic}
\centering
\begin{tabularx}{\columnwidth}{||l||X|X|X||}
\hline\hline
&\multicolumn{3}{c||}{\textbf{Memory access patterns}} \\
\cline{2-4}
\textbf{Platform}    
& \textbf{No reads (TA invocations only)}
& \textbf{One read per 4~kB page}
& \textbf{Read every byte}
\\
\hline \hline

\KinibiTZ
& \begin{tabular}[c]{@{}l@{}}55.84 ms\\ ($\pm$6.04)\end{tabular}
& \begin{tabular}[c]{@{}l@{}}159.06 ms\\ ($\pm$34.52)\end{tabular}
& \begin{tabular}[c]{@{}l@{}}234.44 ms\\ ($\pm$ 55.94)\end{tabular}  
\\
\hline

Intel SGX
 & \begin{tabular}[c]{@{}l@{}}0.37 ms\\ ($\pm$0.02)\end{tabular}
 &  \begin{tabular}[c]{@{}l@{}}0.67 ms\\ ($\pm$0.03)\end{tabular}
 &  \begin{tabular}[c]{@{}l@{}}10.32 ms\\ ($\pm$0.59)\end{tabular}  
 \\ \hline \hline

\end{tabularx}
\end{table}

As shown in Table~\ref{tbl:statistic}, we experimentally measured the average time to cycle a 116~MB dictionary representation ($Y$) through TA in 1~MB chunks, for different memory access patterns.
The first column shows the time required to perform 116 TA invocations without any memory access or computation.
As confirmed by this column, a main strength of Intel SGX is that its enclave entries/exits add very little overhead. 
The second column shows the time for accessing one byte per 4~KB page in $Y$, in addition to TA invocations.
The third column shows the total time for accessing the entirety of $Y$, also in addition to TA invocations.
All read operations were performed using the maximum register size on each platform (i.e. 32~bit on \KinibiTZ{} and 64~bit on Intel SGX).
For Intel SGX, TA invocation overhead is negligible, so overhead shown in the last column is almost entirely due to the read operations.
We can see that more memory accesses result in longer carousel cycling time for both platforms.  
However, even if a dictionary representation allows otherwise, we always access the entirety of $Y$ to ensure query privacy.
Therefore, the last column represents a lower bound for carousel cycling time (and hence query response latency).

\raggedbottom

\pagebreak

\section{Code Optimization}
\label{app:code_optimization}

Implementing algorithms from Section~\ref{sec:dictRep} naively does not ensure that the TA performs equal number of operations on every item in $Y$ at machine-level instructions.
For example, in Algorithm~\ref{algo:cuckoo}, \texttt{R} can be an \texttt{unsigned char} array and \texttt{dummy\_byte} an \texttt{unsigned char} variable. The compiler uses different sets of instructions to copy values of $Y$ on to them causing unequal number of machine-level instructions at the conditional clauses (\texttt{if} and \texttt{else}).
Similarly, the compiler removes or optimizes the dummy operation (e.g. \texttt{dummy\_int ++}) if they are not used elsewhere in the code. It also removes dummy conditional clauses that are unreachable~/~unnecessary.

We tailored our implementation to achieve a balanced set of instructions for the conditional clauses encountered while processing the carousel. Figure~\ref{Ccode} depicts a section of the carousel processing code for Cuckoo hash method that produces equal number of operations on every item in $Y$ at machine-level instructions. 
Figure~\ref{ASMcode} shows the disassembled machine-level instructions mnemonics for the same code segment.
For simplicity the code segment shown in the figure is for processing 16-bit ($\varepsilon=14$) items in $Y$.

In Figure~\ref{Ccode}, \texttt{ptr\_query\_rep} represents the pointer to $S$. 
We use the same variable to represent the dictionary positions as well as store the value of the corresponding position. We implemented the code to operate on 32-bit values. The variables \texttt{ptr\_query\_rep}, \texttt{ptr\_chunk} and \texttt{ptr\_chunk\_end} are defined as \texttt{unsigned int*}. Similarly \texttt{dummy\_pos} is an array of type \texttt{unsigned int}.




\begin{figure*}[ht]
\centering
\begin{lstlisting}[frame=single]

  // ptr_chunk: pointer to the 
  // begining of Y chunk

  // ptr_chunk: pointer to the
  // end Y chunk

  // y_pos: current position
  // in Y

  // ptr_query_rep: pointer to S

  // dummy_pos: dummy array of size 255

 while(ptr_chunk < ptr_chunk_end)
 {
    if(y_pos == *ptr_query_rep)
    {
      *ptr_query_rep = *ptr_chunk;
      ptr_query_rep++;
    } else {
      dummy_pos[(uint8_t)*ptr_chunk] = \
      *ptr_chunk;    
    }
    y_pos ++;
    ptr_chunk = ptr_chunk + 1;
}

\end{lstlisting}
\caption{Kinibi TA code for Cuckoo-on-a-Carousel processing}
\label{Ccode}
\end{figure*}

\begin{figure*}[ht]
\begin{lstlisting}[frame=single]

70e:  1b61       subs    r1, r4, r5
710:  4439       add     r1, r7
712:  f5b1 1f40  cmp.w   r1, #3145728  ; 0x300000
716:  f1c5 0200  rsb r2, r5, #0
71a:  d20a       bcs.n   732 <tlMain+0x1b4>
71c:  6819       ldr r1, [r3, #0]
71e:  4422       add r2, r4
720:  5dd2       ldrb    r2, [r2, r7]
722:  428f       cmp r7, r1
724:  bf0c       ite  eq
726:  f843 2b04  streq.w r2, [r3], #4
72a:  f84a 2022  strne.w r2, [sl, r2, lsl #2]
72e:  3701       adds  r7, #1
730:  e7ed       b.n   70e <tlMain+0x190>

\end{lstlisting}
\caption{Disassembled machine instructions mnemonics for Cuckoo-on-a-Carousel processing}
\label{ASMcode}
\end{figure*}

\end{document}